\documentclass{aa}

\usepackage{graphicx}
\usepackage{comment}
\usepackage{txfonts}

\usepackage[colorlinks=true, allcolors=blue]{hyperref}
\makeatletter
\renewcommand*\aa@pageof{, page \thepage{} of \pageref*{LastPage}}
\makeatother

\begin{document} 

   \title{Predictions for the X-ray polarisation modulation in Cygnus X-1 from reflection off the stellar companion and its wind}

   \titlerunning{X-ray polarisation from reflection in Cyg X-1}

   \authorrunning{B.\ Vander Meulen et al.}

   \author{Bert Vander Meulen
          \inst{1, 2}
          \and
          Kun Hu
          \inst{3}
          \and
          Victoria Grinberg
          \inst{1}
          \and
          Henric Krawczynski
          \inst{4}
          }

   \institute{European Space Agency (ESA), European Space Research and Technology Centre (ESTEC), Keplerlaan 1, 2201 AZ Noordwijk, The Netherlands\\
    \email{bert.vandermeulen@esa.int}
    \and
    Department of Physics and Astronomy, Universiteit Gent, Proeftuinstraat 86 N3, B-9000 Ghent, Belgium
    \and
    Washington University in St. Louis, Physics Department, McDonnell Center for the Space Sciences, Brookings Dr. 1, 63130 St. Louis, MO, USA
    \and
    Washington University in St. Louis, Physics Department, McDonnell Center for the Space Sciences, and Center for Quantum Leaps, Brookings Dr. 1, 63130 St. Louis, MO, USA
    }
   \date{Received \today; accepted ...}


  \abstract
   {Cygnus X-1 is one of the brightest X-ray binaries, and has been observed multiple times with the Imaging X-ray Polarimetry Explorer (IXPE). Recent studies report tentative evidence for a polarisation modulation with the orbital period $P_{\rm orb}$, but a half-period ($P_{\rm orb}/2$) signal, expected from reflection off the companion star and its stellar wind, has not been reported.}
   {We aim to quantify the reflection-induced variations of the polarisation degree PD and polarisation angle PA in Cygnus X-1 as a function of orbital phase and photon energy, and interpret these in terms of binary geometry and wind structure.}
   {We set up a radiative transfer model combining a general relativistic description of the polarised X-ray source emission (kerrC) with a focussed stellar wind model for the binary medium. Using the 3D X-ray radiative transfer code SKIRT, we simulate broadband Stokes I, Q, and U fluxes, surface brightness maps, and linear polarisation maps over one binary orbit.}
   {We find a prominent double-peaked ($P_\mathrm{orb}/2$) polarisation modulation, with a peak-to-peak PD amplitude of $0.25$, $0.81$, and $1.24$ percentage points in the $2-4$, $4-6$, and $6-8~\text{keV}$ bands, respectively, with a strong energy dependence. The PA modulation is more modest, with $|\Delta\text{PA}| < 4.6^\circ$. Crucially, X-ray reprocessing reduces the overall PD relative to the intrinsic source polarisation.}
   {The polarisation modulation is driven by reflection off the companion star and the focussed stellar wind, which induces a polarisation signal that alternately reinforces and counteracts the source polarisation throughout the orbit. The diffuse scattering halo surrounding the X-ray source drives no modulation, but systematically reduces the PD through its anisotropic illumination by the disk-corona system, an effect that should be accounted for in all wind-fed X-ray binaries. The PD amplitude increases with energy as photo-absorption disproportionately attenuates the distant-reflection signal; as the extinction drops, the reflection signal becomes increasingly important.}


   \maketitle
%
\section{Introduction}
\label{sec:intro}
The launch of the Imaging X-ray Polarimetry Explorer (IXPE) mission in 2021 has opened X‑ray polarimetry as a new window on the high-energy Universe, enabling precise measurements of the linear polarisation degree PD and electric-field polarisation angle PA in the $2-8~\text{keV}$ band, with PD sensitivities below $1\%$ for bright Galactic and extragalactic sources \citep{weisskopf22}. Among IXPE’s most prominent targets are black hole X‑ray binaries (BHXRBs), for which significant polarisation has now been detected across hard, soft, and intermediate accretion states \citep[see][for recent summaries of the observational IXPE results]{cavero23, Dovciak24, majumder26}.

Cygnus X-1 (Cyg X-1) was the first BHXRB to be observed with IXPE \citep{krawczynski22b}, being one of the brightest and best-studied X-ray sources in the sky \citep{Jiang_2024}. Cyg X-1 is a binary system consisting of a 21.2$^{+2.2}_{-2.3}\,\text{M}_{\odot}$ black hole and a $40.6^{+7.7}_{-7.1}\,\text{M}_{\odot}$ O9.7\,Iab supergiant companion star, HDE~226868, in a 5.6-day binary orbit viewed at $i={27.5^{\circ}}^{+0.8^{\circ}}_{-0.6^{\circ}}$, at a distance of $2.22^{+0.18}_{-0.17}\,\text{kpc}$ \citep[][but see \citealt{ramachandran25}]{millerjones21}. The X-ray emission is powered by accretion of the strong stellar wind from the companion onto the black hole, and the wind structure can be probed through the phase-dependent absorption variability \citep[e.g.][]{Balucinska-Church_2000a, Boroson_2010a, grinberg15, Yamada_2025}. 

Cyg X-1 has since been observed multiple times with IXPE, and PD and PA have been measured with good accuracy in the hard \citep{krawczynski22b} and soft \citep{steiner24} states, with $2-8~\text{keV}$ polarisation degrees between $2\%$ and $4\%$, and a polarisation direction that is aligned with the radio-jet axis. In both states, PD increases with energy across the IXPE band, while PA shows no significant energy dependence. The hard-state polarisation is consistent with Comptonisation in a hot corona extended in the accretion-disk plane, and suggests that the inner disk is viewed at a substantially higher inclination than the binary orbit. In the soft state, PD is a factor of two lower than the hard-state value, and polarisation may be attributed to returning radiation reflected off the accretion disk, constraining the black hole spin to $a \gtrsim 0.96$ (\citealt{krawczynski22b, steiner24}, but alternative interpretations are discussed in \citealt{beloborodov98, poutanen23, tomaru24, krawczynski25, niedzwiecki26}).

Recently, \citet{kravtsov25} searched for time variability in the Cyg X-1 polarisation properties, scrutinising PD and PA for orbital variations with $P_{\rm orb}$ (i.e., 5.6~\text{days}; \citealt[][]{2008ApJ...678.1237G}) and longer-term, superorbital variability. The authors find tentative evidence for a 5.6-day modulation, which interestingly does not match the $P_{\rm orb}/2$ modulation expected from reflection off the companion star and its wind \citep[][]{brown78, kravtsov20, kravtsov25, rankin24, ahlberg24}. However, this $P_{\rm orb}/2$ signal may still be present at a level inaccessible with current observations, or could be masked by accretion-state-dependent polarisation changes.

This motivates a quantitative study of the reflection-induced PD and PA variations in Cyg X-1. In this work, we model X-ray reprocessing in the 3D binary system using the Monte Carlo radiative transfer code SKIRT \citep{vandermeulen23, vandermeulen24b}, combining a general relativistic model for the polarised X-ray source emission \citep[kerrC;][]{krawczynski22a} with a `focussed' stellar wind model that captures the essential geometry of the Cyg X-1 system \citep{gies86a, grinberg15}; in future work, this model will be refined by incorporating clumping and a more realistic asymmetric wind structure \citep{grinberg15, miskovicova2016}. We predict how reflection off the companion star and its wind modulates PD and PA over one orbital period, and interpret the phase- and energy-dependence of the polarisation signal in terms of the system geometry. In a forthcoming paper, these predictions are confronted with the accumulated IXPE data.

The outline for this paper is as follows: In Sect.~\ref{sec:model}, we introduce the radiative transfer model, which comprises the kerrC source model (Sect.~\ref{sec:kerrC}) and the binary system geometry (Sect.~\ref{sec:focussedwind}). The SKIRT implementation is described in Sect.~\ref{sec:RTsetup}, and the radiative transfer results are presented and discussed in Sect.~\ref{sec:Results}. Our findings are summarised in Sect.~\ref{sec:summary}.

\section{Radiative transfer model}
\label{sec:model}
This work combines a general relativistic model for the polarised X-ray source emission from the disk-corona system \citep[kerrC;][]{krawczynski22a} with a geometric model for the binary system, comprising the companion star HDE~226868 and its focussed stellar wind \citep{grinberg15}. Both components are introduced in the following two subsections; their implementation in SKIRT is discussed in Sect.~\ref{sec:RTsetup}.

\subsection{kerrC polarised source emission}
\label{sec:kerrC}
\begin{figure}[t]
    \centering
	\includegraphics[trim={13 13 8 10}, clip, width=\columnwidth]{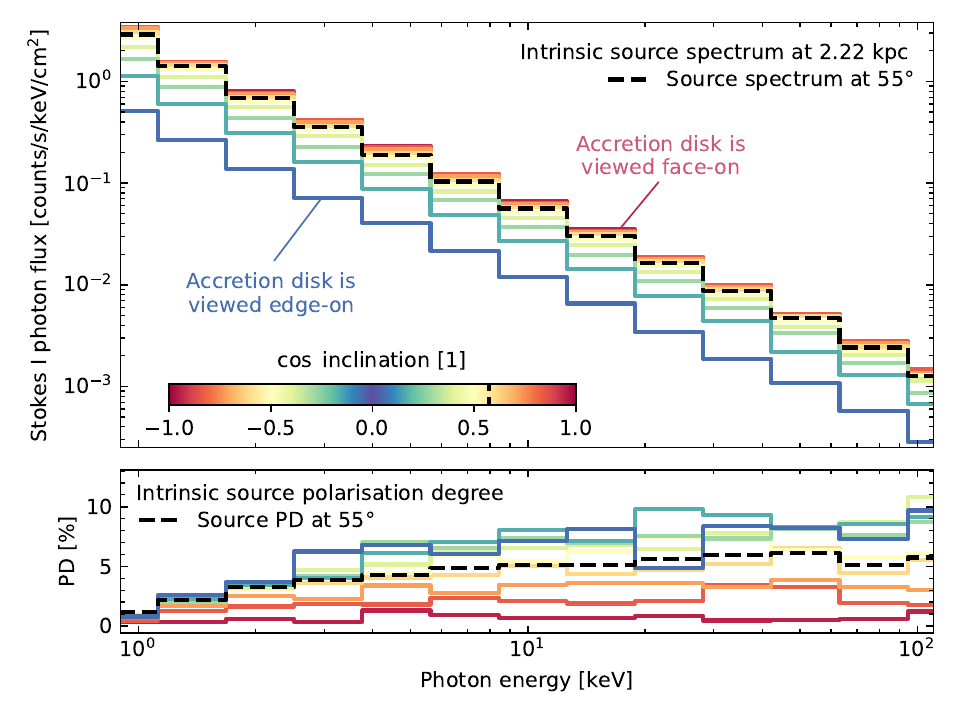}
    \caption{Inclination-dependent polarised source emission from kerrC. Top: Stokes~I photon flux spectrum for different inclinations relative to the accretion-disk axis. Bottom: Corresponding polarisation degree PD; the polarisation direction is always aligned with the projected accretion-disk axis (defined as $\text{PA}=0^\circ$ in this work; Sect.~\ref{sec:roll}). The direct source emission at $55^\circ$ (black) corresponds to $i=27.5^\circ$ relative to the binary orbit (see text). The emission pattern is symmetric about the disk plane.}
    \label{fig:kerrC}
\end{figure}
kerrC is a general relativistic ray-tracing model to simulate and fit the X-ray flux and polarisation spectra (i.e., Stokes I, Q, and U) of accreting black hole systems \citep{krawczynski22a}. The model accounts for thermal emission emitted from a multi-temperature accretion disk and the subsequent Comptonisation within a hot corona. It also self-consistently includes the returning and reflected radiation. X-ray photons are ray-traced from the disk to a distant observer using a Kerr metric, and polarisation vectors are parallel transported along the photon geodesics. For a given set of black hole and corona parameters, kerrC calculates the flux and polarisation spectra for all viewing angles, which can be used to fit spectro-polarimetric observations of BHXRBs \citep[see e.g.][]{krawczynski22a, krawczynski22b, west23, cavero23, marra24, steiner24, krawczynski24, awaki25}.

The adopted kerrC model for the primary X-ray emission of Cyg X-1 corresponds to a thin accretion disk surrounding a black hole with spin parameter $a=0.95$, sandwiched by a wedge-shaped corona with a temperature of $kT=100~\text{keV}$. We assume an opening angle of $10^\circ$ for the corona and set the vertical optical depth to $\tau=0.5$; the mass accretion rate is set to $3.5\times 10^{13}~\text{kg s}^{-1}$. For these parameters, the kerrC flux spectrum at $55^\circ$ (i.e., relative to the accretion-disk axis) matches Cyg X-1 in its hard state. Moreover, the corresponding polarisation is consistent with Cyg X-1 measurements from IXPE, finding a PD increasing from $4$ to $6\%$ in the $2-8~\text{keV}$ band, and {\it XL-Calibur}, finding a PD of about 5\% in the $19-64~\text{keV}$ band \citep{krawczynski22b, awaki25}, with a polarisation angle in the direction of the projected accretion-disk axis (defined as $\text{PA}=0^\circ$ throughout this work; Sect.~\ref{sec:roll}). Our kerrC model for the polarised X-ray source emission of Cyg X-1 is shown in Fig.~\ref{fig:kerrC}.

The high PD measured by IXPE implies that the accretion disk is observed at a high viewing angle ($\sim$$55^\circ$), which is substantially more edge-on than the binary orbit \citep[$i=27.5^\circ$;][]{millerjones21}. This discrepancy could be reconciled by either invoking a relativistic outflowing corona, which would increase the polarisation degree at lower viewing angles \citep{beloborodov98, poutanen23}, or by considering a misalignment between the accretion disk and the binary orbit \citep{krawczynski22b}. The latter scenario is adopted in this work, but we stress that the polarisation variations induced by reflection within the binary system depend only weakly on the source model. We assume that the accretion disk is observed at a viewing angle of $55^\circ$ (Fig.~\ref{fig:kerrC}, black line), consistent with the best-fit kerrC model and recent XRISM results \citep{draghis25}. This implies a misalignment of $\Delta i=27.5^\circ$ between the accretion disk and the binary orbit (for which we adopt $i=27.5^\circ$).

\subsection{Binary geometry with focussed wind}
\label{sec:focussedwind}
\begin{figure}
    \centering
	\includegraphics[width=\columnwidth]{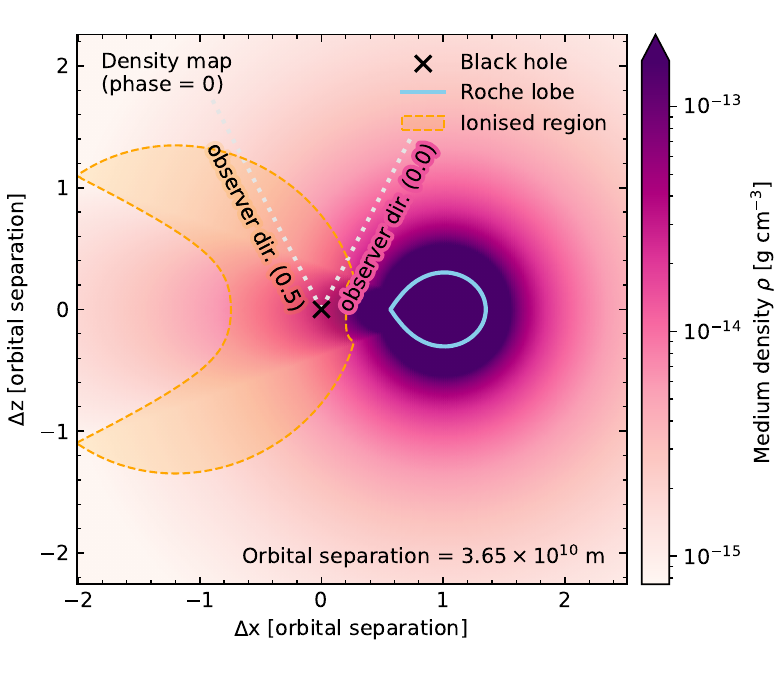}
    \caption{Geometrical model for the Cyg X-1 binary system, comprising the black-hole X-ray source, the companion star, and a focussed stellar wind. The $x$-axis is the cylindrical symmetry axis connecting the black hole (centred on the origin) and the companion star (located at $x=1$ in units of orbital separation). The ionised wind region is indicated in orange (see text). The binary system is observed at $i = 27.5^\circ$ (Sect.~\ref{sec:kerrC}); the observer directions at phases $0.0$ and $0.5$ are indicated in white.}
    \label{fig:focussedwind}
\end{figure}
Before reaching the observer, the primary X-ray emission from the inner accretion flow (Sect.~\ref{sec:kerrC}) is reprocessed within the binary system, by the supergiant companion star HDE~226868 and its wind. The optically thick photosphere of the companion provides a large reflective surface, whose illuminated face rotates in and out of the line of sight over the binary orbit. The companion also drives a strong stellar wind \citep{castor75}, which permeates the system and constitutes an additional, extended scattering medium.

As HDE~226868 almost fills its Roche lobe \citep{Gies_2003a, ramachandran25}, the tidal forces exerted by the black hole focus the stellar wind towards the black hole, breaking the spherical symmetry and forming a density enhancement along the donor-black hole axis \citep{gies86a, gies86b, Sowers_1998a, Miller_2005a, Hanke_2009a, El_Mellah_2019a, brigitte25}. We model this with a smooth, axisymmetric wind model consisting of a spherical Castor-Abbott-Klein component \citep[CAK;][]{castor75} and a focussed wind stream directed towards the companion star \citep{friend82, gies86a, gies86b}, as shown in Fig.~\ref{fig:focussedwind}. Following \citet{gies86a}, the focussed component occupies a conical region with an opening angle of $20^\circ$, where the wind is denser and slower than the CAK solution; outside this cone, the standard single-star CAK solution applies \citep[see also][]{miskovicova2016}. This model provides a good representation of the non-clumped wind in Cyg X-1 \citep{grinberg15,miskovicova2016}, reproducing the observed optical \citep{gies86a} and X-ray variability \citep{grinberg15, Hirsch_2019a}; in future work it will be refined to incorporate clumping \citep[e.g.][]{Owocki_1984a, Sundqvist_2018a} and orbital deflections of the accretion stream \citep[e.g.][]{El_Mellah_2019a}.

We adopt the system parameters from \citet{millerjones21}, updating the focussed-wind model employed in \citet{grinberg15} \citep[see also][]{lai24}. Since the \citet{gies86a} wind model scales with system size, we obtain the same density distribution in units of orbital separation. However, as the updated separation of $3.65\times10^{10}~\text{m}$ exceeds the separation adopted in \citet{grinberg15} by a factor of $1.24$, we rescale the volume density by $1/1.24$ to preserve the observed $N_\text{H}$ trend as a function of orbital phase. This density scaling is well within the uncertainties on the underlying mass-loss rate and wind velocity structure \citep{ramachandran25}. We note that the same procedure applies if alternative parameters are adopted: the density rescaling naturally recovers identical radiative transfer results, making our approach robust to ongoing revisions of the Cyg X-1 system parameters \citep[e.g.][]{ramachandran25}.

The 2D density map of the binary system is shown in Fig.~\ref{fig:focussedwind}, representing the stellar companion and its wind. To obtain the 3D radiative transfer medium, this map must be revolved about the donor-black hole axis (i.e., the $x$-axis), which is the cylindrical symmetry axis of the system. Following \citet{grinberg15}, we define two regions within the wind, assuming that Cyg X-1 is in the hard state: a highly ionised region close to the black hole, where the wind contributes to reflection but not absorption; and a more neutral region further away that contributes to both absorption and reflection. For details on the calculation of the ionised regions, readers can refer to \citet{grinberg15}. Similarly, we rescale the threshold ionisation parameter $\xi_\text{max}$ of \citet{grinberg15} by $1/1.24$, to preserve the ionisation boundary in units of orbital separation, and thereby the observed absorption trend with orbital phase. The highly ionised wind region, corresponding to $\log \xi \geq 2.62$, is marked in orange on Fig.~\ref{fig:focussedwind}.

\section{Radiative transfer setup}
\label{sec:RTsetup}
We employ the Monte Carlo radiative transfer code SKIRT\footnote{We use the public master branch of SKIRT v9, git commit \texttt{d762d4b}, publicly available at: \url{https://skirt.ugent.be}.}
\citep{camps15a, camps20, vandermeulen23, vandermeulen24b} to propagate polarised kerrC source photons through the Cyg X-1 binary system, and predict the X-ray polarisation signal affected by reflection off the companion star and its wind. SKIRT models X-ray reprocessing in 3D, emulating photo-absorption with self-consistent fluorescent line emission and electron scattering, with full support for X-ray polarisation. For technical details, readers can refer to \citet{vandermeulen23, vandermeulen24b}.

In this study, two novel SKIRT features are deployed, which are introduced in Sect.~\ref{sec:FPPS} and Sect.~\ref{sec:CCG}. The general radiative transfer setup is described in Sect.~\ref{sec:generalsetup}.

\subsection{kerrC source emission in SKIRT}
\label{sec:FPPS}
The general-relativistic kerrC model (Sect.~\ref{sec:kerrC}) provides the inclination-dependent flux and polarisation spectra of the X-ray source in Cyg X-1 (i.e., the inclination- and energy-dependent Stokes I, Q, and U parameters). To import these kerrC results to the SKIRT code, we introduced a new SKIRT function to model sub-grid source emission with an advanced emission pattern: the \texttt{FilePolarizedPointSource}\footnote{\url{https://skirt.ugent.be/skirt9/class\_file\_polarized\_point\_source.html}}.

The \texttt{FilePolarizedPointSource} class represents a single point source, emitting polarised radiation with an axisymmetric angular dependence, based on user-provided Stokes spectra that are tabulated as a function of inclination cosine. For this project, relevant configurable options are the location of the point source and the direction of the symmetry axis.

The kerrC Stokes spectra are provided for 18 linear cosine bins between $-1$ and $1$, tabulated on a logarithmic energy grid with $E/\Delta E = 2.5$, covering the $1-200~\text{keV}$ range. These tables are interpolated by the \texttt{FilePolarizedPointSource} to obtain normalised Stokes parameters for each direction and energy. The source is located at the origin, and the source symmetry axis is tilted by $\Delta i=-27.5^\circ$ in the azimuthal plane, such that the accretion-disk axis is always observed at $55^\circ$ when the binary system is observed at $i = 27.5^\circ$, as discussed in Sect.~\ref{sec:kerrC}.


\subsection{Focussed-wind transfer medium in SKIRT}
\label{sec:CCG}
The SKIRT code has various interfaces for importing tabulated transfer media \citep[see e.g.][for recent examples]{baes24a, baes24b, baes25, gebek24, gebek25, matsumoto24, matsumoto25, kapoor24}. In this work, we have a 2D model representing the axisymmetric 3D density distribution of the Cyg X-1 system (Sect.~\ref{sec:focussedwind}), tabulated on a Cartesian grid in the azimuthal plane. This can be imported as a transfer medium in SKIRT using the \texttt{CylindricalCellGeometry} class\footnote{\url{https://skirt.ugent.be/skirt9/class\_cylindrical\_cell\_geometry.html}}, which has the capability to `revolve' a tabulated 2D density distribution around a cylindrical symmetry axis (here: the $x$-axis) to obtain a 3D transfer medium.

The reason for employing the \texttt{CylindricalCellGeometry}, which only loads the (normalised) 3D geometry, instead of the \texttt{CylindricalCellMedium}, which imports the gas mass density in physical units, is that the latter cannot be rotated. In SKIRT, cylindrical imports always place the cylindrical symmetry axis along the $z$-axis. However, this $z$-axis also defines the inclination angle of the observer and the reference direction for polarisation. Without further rotation, this would complicate our analysis: (i) the inclination in SKIRT would not correspond to the inclination of the binary orbit, and (ii) recorded Stokes vectors would not be defined relative to the projected orbital axis. This is resolved by combining the \texttt{Cylindrical\allowbreak Cell\allowbreak Geometry} with a \texttt{Rotate\allowbreak Geometry\allowbreak Decorator}, with $\texttt{eulerBeta}=90^\circ$, which rotates the medium back to the $x-y$ plane, and $\texttt{eulerGamma}=90^\circ$, which realigns the companion with the positive $x$-axis (Fig.~\ref{fig:focussedwind}). 

To obtain the gas mass normalisation for the (normalised) \texttt{CylindricalCellGeometry}, we run a single simulation with the corresponding \texttt{CylindricalCellMedium}, which loads the gas mass density in physical units; the resulting normalisation is extracted from the convergence information file and applied to the (rotated) \texttt{CylindricalCellGeometry} component.

X-ray interactions in the binary medium are implemented in the \texttt{XRayAtomicGasMix}, which models absorption, scattering, and fluorescent re-emission by cold gas with X-ray polarisation support. We adopt \citet{wilms00} abundances, with a total gas mass of $2.35 \times 10^{-24}~\text{g}$ per H-atom, consistent with previous studies of Cyg X-1 \citep{grinberg15, lai24}.

Closer to the X-ray source, we assume that the wind is fully ionised, as discussed in Sect.~\ref{sec:focussedwind} \citep[see also][]{grinberg15}. In the ionised region (marked in orange on Fig.~\ref{fig:focussedwind}) we adopt the \texttt{ElectronMix} with $1.20$ free electrons per H-atom, modelling electron scattering with polarisation support, without absorption or fluorescent re-emission.

\subsection{General radiative transfer setup}
\label{sec:generalsetup}
The \texttt{RotateGeometryDecorator} mentioned in Sect.~\ref{sec:CCG} could have actively rotated the binary medium as a function of orbital phase\footnote{Varying $\texttt{eulerGamma}$ as $(90^\circ + 360^\circ \times \texttt{phase})\pmod{360^\circ}$, which corresponds to the standard phase convention of \citet{gies86a}, with the orbit appearing clockwise on the sky \citep{millerjones21}. We note that the rotation decorator implements a passive coordinate transformation, equivalent to an active rotation by the opposite angle.}. Instead, we adopt the co-rotating binary frame, and vary the observer direction and source symmetry axis, which allows the SKIRT medium to be discretised on a spatial grid that closely follows the tabulated focussed-wind model (Sect.~\ref{sec:focussedwind}). We adopt a 3D Cartesian grid covering
$-9.15\times 10^{10}$ to $7.33\times 10^{10}~\text{m}$ in 300 steps in the $x$-direction and $-8.24\times 10^{10}$ to $8.24\times 10^{10}~\text{m}$ in 300 steps in the $y$- and $z$-directions, which corresponds to a uniform spatial resolution of $5.49\times 10^{8}~\text{m}$ in each direction.

Simulation results are recorded for an observer at $2.22~\text{kpc}$, with $i=27.5^\circ$ \citep{millerjones21} and an azimuthal angle of $(360^\circ \times \texttt{phase})\pmod{360^\circ}$ in the co-rotating frame. This corresponds to the phase convention of \citet{gies86a}, where phase zero is defined as the inferior conjunction of the companion star, and to an orbital angular momentum vector pointing away from the observer (i.e., clockwise orbital motion on the sky; \citealt{millerjones21}). We run a total number of 41 SKIRT simulations, covering one full orbit in 41 phase steps of $0.025$. We note that a single simulation with multiple observers cannot provide all radiative transfer results at once, as the relative orientation of the source symmetry axis and the binary medium varies as a function of orbital phase (Sects.~\ref{sec:kerrC} and~\ref{sec:FPPS}).

Integrated fluxes and surface brightness maps are produced in three energy bands ($2-4~\text{keV}$, $4-6~\text{keV}$, and $6-8~\text{keV}$), while spectra are recorded on a logarithmic energy grid that covers the $1 - 100~\text{keV}$ range in 500 bins ($E/\Delta E = 108$). Surface brightness maps are centred on the primary X-ray source, and cover a field of view of $10^{11}~\text{m} \times 10^{11}~\text{m}$ in the sky-projected Cyg X-1 frame. All SKIRT instruments (i.e., broadband, spectral, and imaging) can separate the individual flux contributions of the direct source emission, the electron-scattered continuum, and the fluorescent lines, which provides valuable information for the analysis of the radiative transfer results (Sect.~\ref{sec:Results}). 

The X-ray source emission and all radiative processes are modelled up to $200~\text{keV}$, as required to achieve fully converged X-ray flux and polarisation spectra up to $\sim$$100~\text{keV}$ \citep[see][for a discussion]{vandermeulen23}. The number of discrete photon packets launched from the kerrC point source is $5\times10^8$, which offers an optimal balance between simulation runtime and Monte Carlo noise.\footnote{These $5\times10^8$ photons do not just escape the model in all directions, with a low probability of being detected, but instead contribute directly to the simulation output at each (re)emission and scattering location through peel-off \citep[see][]{yusefzadeh84}.} Each individual simulation has a runtime of about 35 hours on a single CPU core; on a small 41-core cluster, the full set of 41 simulations completes in roughly 35 hours when run in parallel, totalling about 1,435 core-hours.

\section{Results and discussion}
\label{sec:Results}
\subsection{Radiative transfer results}
\label{sec:RTresults}
\begin{figure*}
    \centering
	\includegraphics[trim={0 0 0 0}, clip,width=\linewidth]{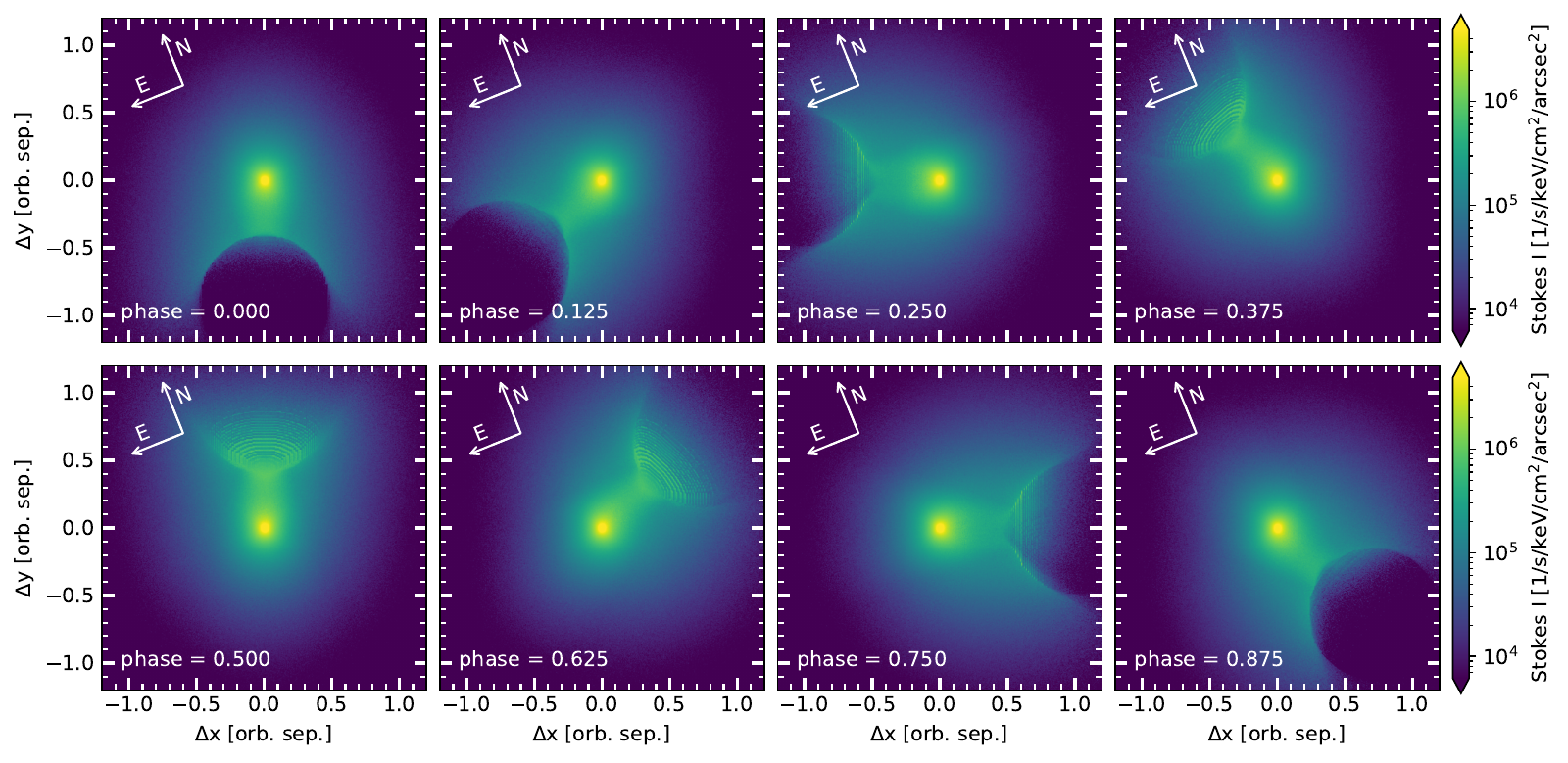}
    \caption{Broadband Stokes I surface brightness maps in the $4-6~\text{keV}$ band at eight equally-spaced orbital phases, providing a direct view of the Cyg X-1 system over one orbit. The binary orbit is observed at $i = 27.5^{\circ}$, with the orbital angular momentum vector pointing away from the observer, and phase zero corresponding to the inferior conjunction of the companion star. The projected accretion-disk axis is aligned with the positive $y$-direction, which corresponds to a rotation of $21^{\circ}$ relative to celestial north (indicated on each panel; Sect.~\ref{sec:roll}). The primary X-ray source is centred at $(0,\,0)$, and the extended emission traces reflection off the companion star and its focussed stellar wind (Sect.~\ref{sec:focussedwind}).}
    \label{fig:images}
\end{figure*}
\begin{figure}
    \centering
	\includegraphics[trim={22 11 41 40}, clip,width=\columnwidth]{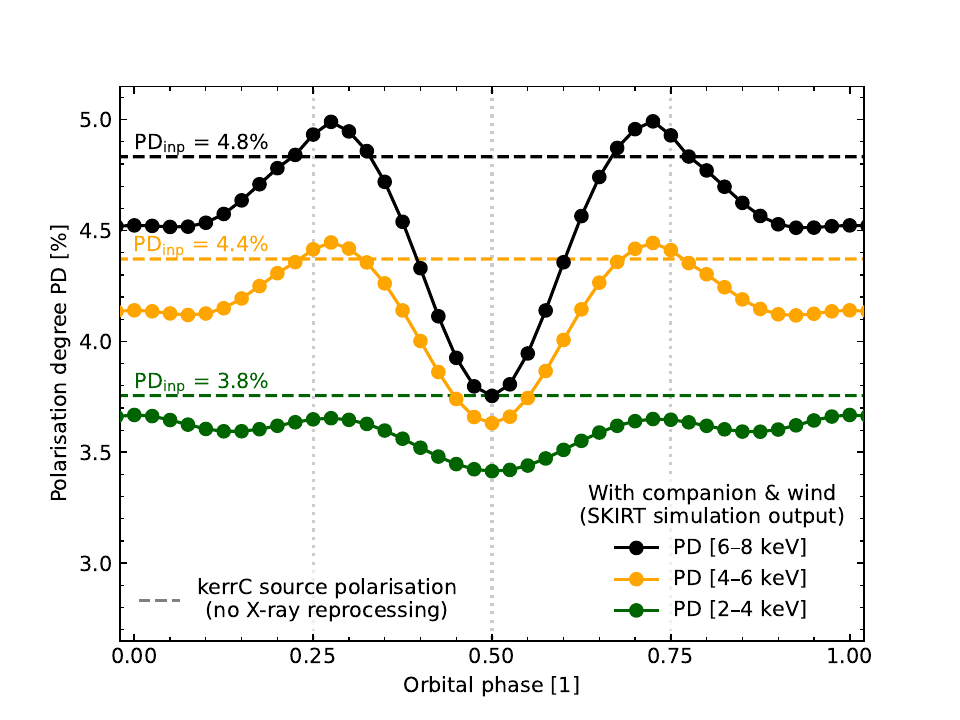}
    \caption{Orbital modulation of the linear polarisation degree PD in three energy bands. The dashed horizontal lines indicate the intrinsic source polarisation in each band ($\text{PD}_\text{inp}$ from kerrC; Fig.~\ref{fig:kerrC}).}
    \label{fig:orbitalModulation}
\end{figure}

Figure~\ref{fig:images} shows the simulated Stokes I surface brightness maps in the $4-6~\text{keV}$ band at eight equally-spaced orbital phases, providing a direct view on the Cyg X-1 system over one orbit. Throughout the orbit, the observed emission is dominated by the direct contribution from the point-like X-ray source (Sect~\ref{sec:kerrC}), while the extended emission traces reflection off the companion star and its stellar wind (Sect~\ref{sec:focussedwind}). The extended emission region rotates about the line of sight with the binary orbit (Fig.~\ref{fig:images}), and contributes most near phase $0.5$, when the focussed wind and the illuminated face of the companion star are maximally exposed. Similar maps were obtained for the $2-4~\text{keV}$ and $6-8~\text{keV}$ bands: at lower energies the diffuse scattered emission dominates, while reflection off the companion star and the focussed stellar wind becomes increasingly important at higher energies.

Our main result is shown in Fig.~\ref{fig:orbitalModulation}: the linear polarisation degree PD as a function of orbital phase, in three energy bands ($2-4~\text{keV}$, green; $4-6~\text{keV}$, orange; $6-8~\text{keV}$, black). We find a prominent double-peaked polarisation modulation, with a peak-to-peak PD amplitude of $0.25$, $0.81$, and $1.24$ percentage points in the three energy bands, respectively, which corresponds to a substantial modulation of the source polarisation degree ($7\%$, $19\%$, and $26\%$). The modulation is symmetric about phase $0.5$, with an energy-dependent amplitude, but otherwise similar phase dependence across all bands; PD is minimal at phase $0.5$, when the companion lies behind the X-ray source, and maximal near phases $0.275$ and $0.725$, when the binary components are roughly side-by-side (Fig.~\ref{fig:images}). The dashed horizontal lines indicate the intrinsic source polarisation in each band ($\text{PD}_\text{inp} = 3.8\%$, $4.4\%$, and $4.8\%$; Fig.~\ref{fig:kerrC}). X-ray reprocessing generally reduces the observed PD, with the exception of a small amplification near phases $0.275$ and $0.725$ in the highest energy bands.

\begin{figure*}
    \centering
	\includegraphics[trim={7 2 6 2}, clip,width=\linewidth]{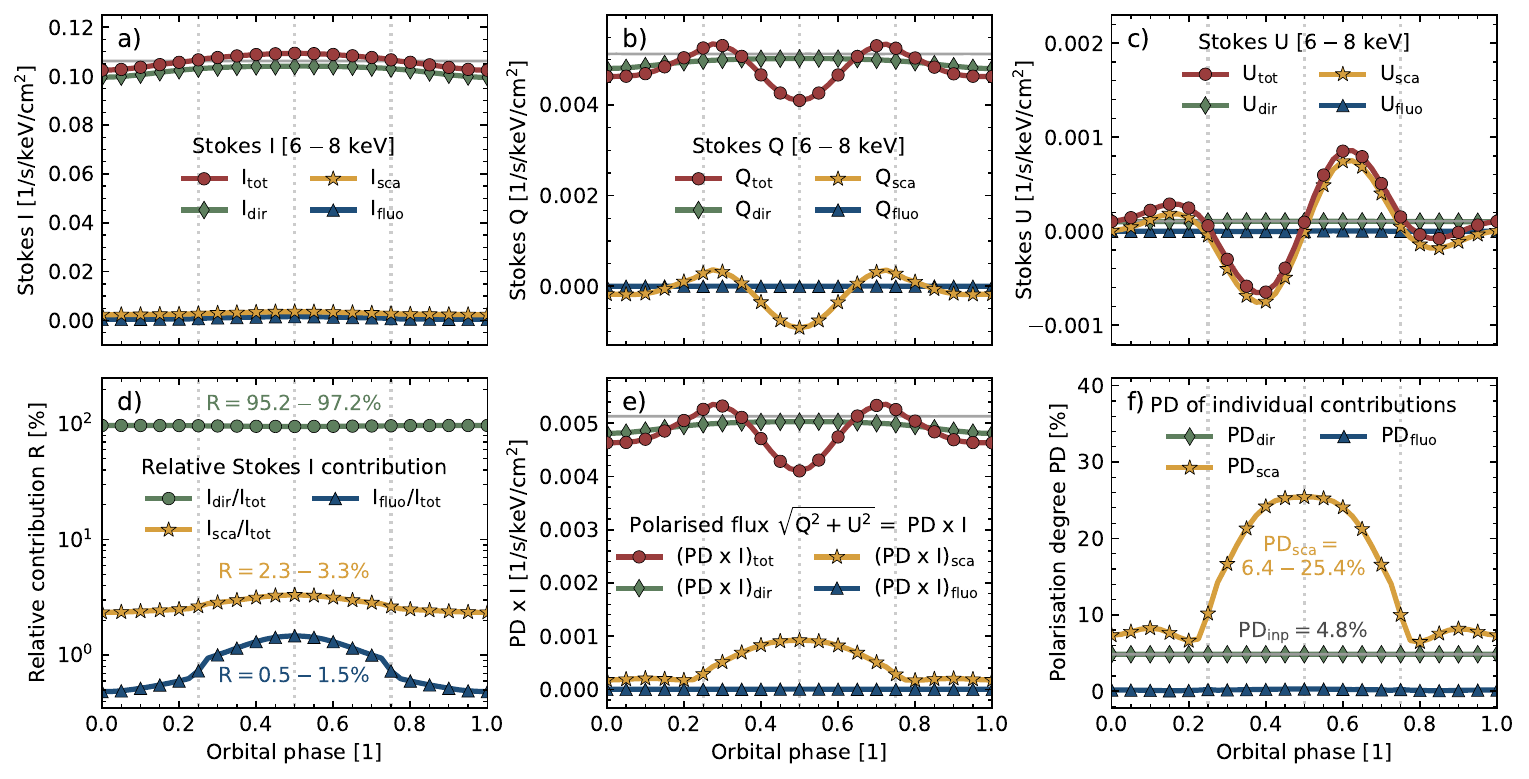}
    \caption{Broadband Stokes I, Q, and U fluxes as a function of orbital phase for the $6-8~\text{keV}$ band. The total observed flux (red) is the sum of three flux contributions: the direct source contribution (green), the Compton-scattered continuum (orange), and the fluorescent-line emission (blue); $\text{I}_\text{tot}=\text{I}_\text{dir}+\text{I}_\text{sca}+\text{I}_\text{fluo}$, and equivalently for Stokes Q and U. The intrinsic primary-source flux is shown in grey as a reference. Stokes Q and U are referred to the projected accretion-disk axis, which is offset from celestial north by $21^{\circ}$ (see Fig.~\ref{fig:images} and Sect.~\ref{sec:roll}).}
    \label{fig:broadband}
\end{figure*}

The linear polarisation degree at each orbital phase (Fig.~\ref{fig:orbitalModulation}) was calculated from the simulated broadband I, Q, and U fluxes, which are shown in Fig.~\ref{fig:broadband} (top row) for the $6-8~\text{keV}$ band, which is the energy band where the PD modulation is most prominent. For each Stokes parameter, the total flux (red) is the sum of three distinct flux contributions: the direct source emission (green), the Compton-scattered continuum (orange), and the fluorescent-line emission (blue); $\text{I}_\text{tot} = \text{I}_\text{dir} + \text{I}_\text{sca} + \text{I}_\text{fluo}$, and equivalently for Q and U. The intrinsic source flux is shown in grey as a reference.

The total Stokes I flux (Fig.~\ref{fig:broadband}a, red) is dominated by the direct source contribution (green), as expected for the modest line-of-sight extinction of Cyg X-1 \citep{grinberg15}. In the $6-8~\text{keV}$ band, the intrinsic source flux is transmitted at $94-98\%$, depending on the orbital phase: the optical depth is minimal at phase $0.5$ ($\tau_\text{ext} = 0.020$), when the companion lies behind the X-ray source, and maximal at phase $0$ ($\tau_\text{ext} =  0.065$), when the line of sight traverses the largest column of stellar wind. The same trend is apparent in $\text{Q}_\text{dir}$ (Fig.~\ref{fig:broadband}b, green), while $\text{U}_\text{dir} \approx 0$ by definition (Sect.~\ref{sec:kerrC}). Since extinction attenuates I and Q by the same factor, both the Q/I ratio and PD of the direct flux contribution remain unchanged at $\text{PD}_\text{inp} = 4.8\%$ throughout the orbit\footnote{This is not generally true, as energy-dependent extinction could alter the flux-weighted broadband PD of the direct contribution. This effect is negligible here ($\left|\Delta \text{PD}\right| < 0.04\%$, $0.004\%$, and $0.0004\%$, respectively), as the source $\text{PD}_\text{inp}$ is roughly constant with energy (Fig.~\ref{fig:kerrC}, bottom).} (Fig.~\ref{fig:broadband}f, green); the direct flux varies, but introduces no polarisation modulation.

A fraction of the source flux absorbed within the binary system is re-emitted as fluorescent line emission (Fig.~\ref{fig:broadband}a, blue). As fluorescent lines are intrinsically unpolarised ($\text{PD}_\text{fluo} = 0$, and hence $\text{Q}_\text{fluo} = \text{U}_\text{fluo} = 0$; Fig.~\ref{fig:broadband}b,c,f), they do not contribute to the polarised flux (Fig.~\ref{fig:broadband}e), and only dilute the total polarisation degree. However, as their relative Stokes I contribution is only $0.5-1.5\%$ (Fig.~\ref{fig:broadband}d), this dilution is negligible, and fluorescent line emission can be ignored in the polarisation analysis.

The remaining flux contribution is the Compton-scattered continuum (Fig.~\ref{fig:broadband}, orange). As neither the direct flux (constant $\text{PD}_\text{dir}$ at $4.8\%$) nor the fluorescent line emission (negligible flux) drives a polarisation modulation, the observed PD modulation must originate entirely from the Compton-scattered continuum, which is discussed in detail in Sect.~\ref{sec:reflection}.

\subsection{Polarisation of the Compton-scattered continuum}
\label{sec:reflection}
\begin{figure*}
    \centering
	\includegraphics[trim={0 0 0 0}, clip,width=\linewidth]{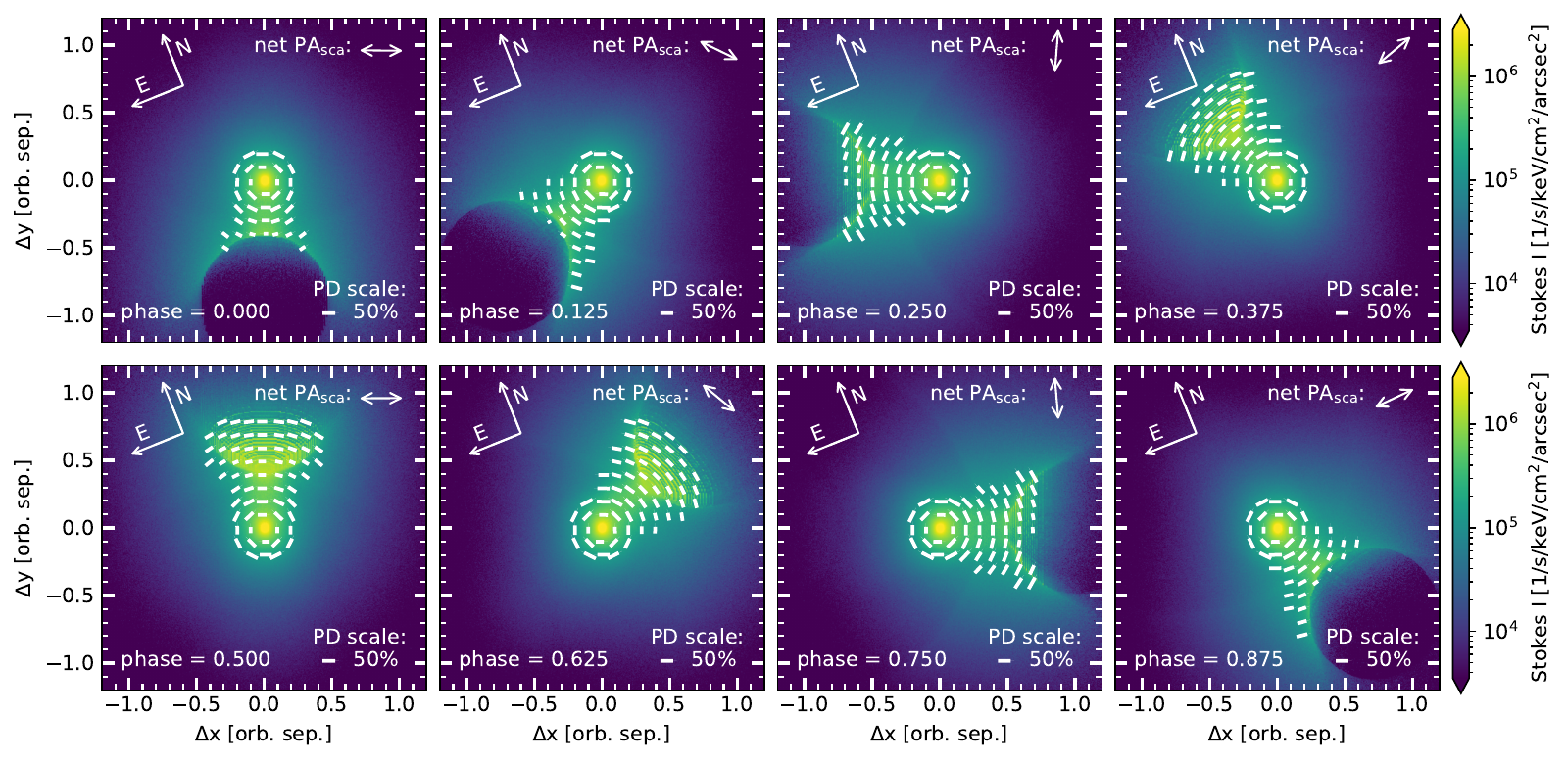}
    \caption{Linear polarisation maps in the $6-8~\text{keV}$ band at eight equally-spaced orbital phases (see Fig.~\ref{fig:images} for more details). The spatially resolved polarisation degree PD and polarisation angle PA are represented by the length and orientation of the white line segments, respectively. The direct source polarisation is aligned with the projected accretion-disk axis (positive $y$-direction), which is offset from celestial north by $21^{\circ}$ (see Fig.~\ref{fig:images} and Sect.~\ref{sec:roll}); the net scattered polarisation direction is indicated on each panel and rotates with orbital phase.}
    \label{fig:polmap}
\end{figure*}
While the Compton-scattered continuum flux contributes only $2.3-3.3\%$ of the total Stokes I flux (Fig.~\ref{fig:broadband}d, orange), this component is highly polarised (up to $25\%$; Fig.~\ref{fig:broadband}f) and therefore contributes notably to the total polarised flux $(\text{PD} \times \text{I}$; Fig.~\ref{fig:broadband}e). Two scattering regions can be identified: (i) a diffuse scattering halo centred on the X-ray source, which is visible throughout the orbit; and (ii) a more distant component from scattering off the companion star and its focussed stellar wind, which dominates when the illuminated face of the companion and the focussed-wind axis are exposed, between phases $0.25$ and $0.75$ (Fig.~\ref{fig:images}).

These two scattering regions contribute differently to the net spatially integrated polarisation. The diffuse stellar wind close to the X-ray source provides a nearly isotropic scattering medium, whose polarisation contributions largely cancel in the integrated flux, yielding a low net $\text{PD}_\text{sca}$ of $6-8\%$ (Fig.~\ref{fig:broadband}f; not exactly zero due to the anisotropic illumination of the scattering medium by the X-ray source, see Fig.~\ref{fig:kerrC}). The companion surface and the focussed wind, by contrast, constitute an anisotropic scattering medium along the donor-black hole axis, inducing significant net polarisation in the spatially integrated flux (up to $25\%$; Fig.~\ref{fig:broadband}f). This is apparent in the polarised flux $(\text{PD} \times \text{I})_\text{sca}$ (Fig.~\ref{fig:broadband}e, orange), which peaks between phases $0.25$ and $0.75$.

The total polarised flux does not only depend on the magnitude of the polarised flux contributions, but also on the relative orientation of their polarisation directions. The polarisation direction of the scattered flux is set by the geometry of the scattering medium: since the scattered polarised flux is dominated by single scattering off the companion star and the focussed wind along the donor-black hole axis, and single scattering induces polarisation perpendicular to the scattering plane \citep[e.g.,][]{vandermeulen24b}, the net scattered polarisation is perpendicular to the sky-projected donor-black hole axis. As this sky-projected axis sweeps about the line of sight throughout the orbit, the scattered polarisation direction rotates with orbital phase.\footnote{At $i=27.5^\circ$, the projected axis does not rotate uniformly: the phase offset relative to uniform motion is zero at phases $0$, $0.25$, $0.5$, and $0.75$, and reaches a maximum offset of $|\Delta\text{phase}| = 0.009$ when the projected binary axis lies diagonally (Fig.~\ref{fig:images}); this projection effect is negligible and does not notably affect the observed polarisation modulation.} This is visualised in Fig.~\ref{fig:polmap}, showing the linear polarisation maps in the $6-8~\text{keV}$ band, derived from the simulated Stokes I, Q, and U surface brightness maps.

The total polarised flux is dominated by the direct source contribution (Sect.~\ref{sec:RTresults}), which is polarised along the projected accretion disk axis (Sect.~\ref{sec:kerrC}; positive $y$-direction in Fig.~\ref{fig:polmap}). As the scattered polarisation direction rotates with orbital phase, it alternately reinforces and counteracts the direct source polarisation, modulating the total polarised flux. At phases $0.25$ and $0.75$, the two polarisation directions align (Fig.~\ref{fig:polmap}), and the scattered flux amplifies the observed polarisation degree (Fig.~\ref{fig:orbitalModulation}). At phases $0$ and $0.5$, they are perpendicular (Fig.~\ref{fig:polmap}), and the scattered flux counteracts the direct source polarisation, reducing the observed PD. This alternating (mis)alignment repeats twice per orbit, naturally producing the double-peaked $P_\mathrm{orb}/2$ polarisation modulation (Fig.~\ref{fig:orbitalModulation}).

The detailed shape of the polarisation modulation is further set by the phase-dependent magnitude of the scattered polarised flux $(\text{PD} \times \text{I})_\text{sca}$ (Fig.~\ref{fig:broadband}e, orange). This peaks near phase $0.5$, when the illuminated face of the companion star is maximally exposed, resulting in a deeper PD minimum at phase $0.5$ than at phase $0$ (Fig.~\ref{fig:orbitalModulation}). Similarly, the PD maxima are shifted from phases $0.25$ and $0.75$ with perfect alignment to phases $0.275$ and $0.725$ where the scattered polarised flux is larger.

\subsection{Visualisation in the Q -- U plane}
\label{sec:QUplane}
\begin{figure*}
\sidecaption
    \centering
	\includegraphics[trim={7 2 6 2}, clip,width=12cm]{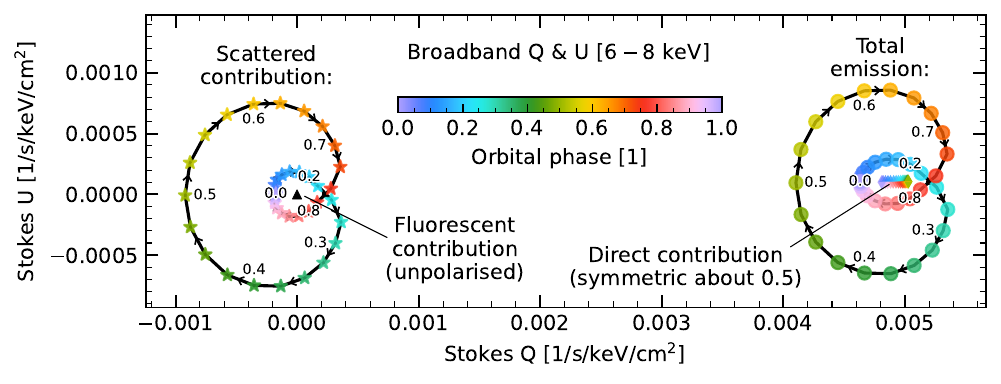}
    \caption{Variation of the broadband Stokes Q and U fluxes in the $6-8~\text{keV}$ band with orbital phase. The total observed flux is the sum of the direct source emission, the reflected emission, and the (unpolarised) fluorescent line emission. The Q-axis is aligned with the primary source polarisation direction (Sect.~\ref{sec:roll}).}
    \label{fig:QU_loop}
\end{figure*}
This interplay between the direct and scattered polarised flux is most clearly visualised in the $\text{Q}-\text{U}$ plane (Fig.~\ref{fig:QU_loop}). The Stokes Q and U parameters describe the polarised flux along two pairs of reference axes, such that orthogonal polarisation directions carry opposite signs and cancel when added: positive Q-values describe polarisation along the sky-projected accretion-disk axis (i.e., the source polarisation direction; Sect.~\ref{sec:roll}), while negative Q-values describe polarisation perpendicular to this axis; Stokes U similarly describes polarisation along orthogonal axes rotated by $45^\circ$ \citep[see e.g.][]{peest17}.

The direct contribution traces a nearly constant positive-Q locus (Fig.~\ref{fig:QU_loop}, diamonds), with a minor symmetric variation about phase $0.5$ due to the variable line-of-sight extinction (Sect.~\ref{sec:RTresults}). The scattered contribution (Fig.~\ref{fig:QU_loop}, stars) rotates twice about the origin throughout the orbit, with the polarised flux distributed sinusoidally between Q$_\text{sca}$ and U$_\text{sca}$ (Fig.~\ref{fig:broadband}b,c); the larger loop corresponds to phases $0.25-0.75$, when the illuminated face of the companion is exposed. The total polarisation signal (Fig.~\ref{fig:QU_loop}, circles) is the vector sum of these two contributions, and reflects the rotation of the scattered polarised flux about the direct source polarisation signal in the $\text{Q}-\text{U}$ plane.

The total polarised flux $(\text{PD} \times \text{I})_\text{tot}$ is the vector norm of ($\text{Q}_\text{tot},\,\text{U}_\text{tot})=(\text{Q}_\text{dir}+\text{Q}_\text{scat},\,\text{U}_\text{scat}$). Since the $\text{U}_\text{sca}$ component is orthogonal to the dominant ($\text{Q}_\text{dir},\,0$) contribution in the $\text{Q}-\text{U}$ plane (Fig.~\ref{fig:QU_loop}), the total polarised flux is to first order insensitive to the $\text{U}_\text{sca}$ signal: $(\text{PD} \times \text{I})_\text{tot} \approx \text{Q}_\text{dir} \times [1 + 2\,\text{Q}_\text{sca}/\text{Q}_\text{dir}]^{1/2}$. The observed PD modulation (Fig.~\ref{fig:orbitalModulation}) therefore closely resembles the orbital trend of $\text{Q}_\text{sca}$ (Fig.~\ref{fig:broadband}b, orange).

\begin{figure}
    \centering
	\includegraphics[trim={22 11 41 40}, clip,width=\columnwidth]{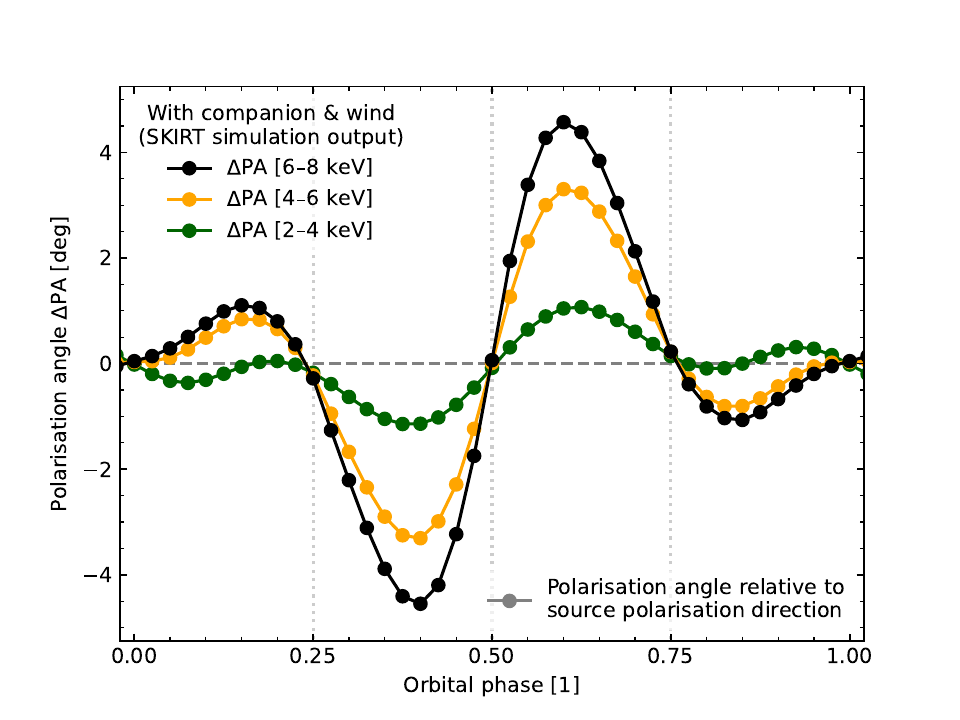}
    \caption{Orbital modulation of the polarisation angle offset, $\Delta\text{PA}$, in three energy bands, relative to the intrinsic source polarisation direction.}
    \label{fig:deltaGamma}
\end{figure}
Similarly, the effect of $\text{U}_\text{sca}$ on the observed net polarisation angle PA is minor, since $\left|\text{U}_\text{sca}\right| \ll \text{Q}_\text{tot}$ (Fig.~\ref{fig:QU_loop}). Still, as $\text{U}_\text{dir} \approx 0$ and $\text{Q}_\text{tot}$ lies along the positive Q-direction, $\text{U}_\text{sca}$ is the only flux contribution that can alter PA, and therefore we expect a minor PA modulation centred on $\text{PA} = 0^\circ$. This can be visualised in the $\text{Q}-\text{U}$ plane (Fig.~\ref{fig:QU_loop}), where PA corresponds to $1/2$ of the angle between ($\text{Q}_\text{tot},\,\text{U}_\text{tot})$ and the positive Q-direction. Fig.~\ref{fig:deltaGamma} shows our predicted modulation of the polarisation angle, with $\left|\Delta\text{PA}\right| < 4.6^\circ$ throughout the orbit, consistent with observational studies reporting no significant variation in PA with orbital phase \citep{krawczynski22b, steiner24, kravtsov25}. Specifically, we find a peak-to-peak PA amplitude of $2.2^\circ$, $6.6^\circ$, and $9.1^\circ$ in the $2-4$, $4-6$, and $6-8~\text{keV}$ bands, respectively. To first order, $\Delta\text{PA} \approx 0.5 \times \text{U}_\text{sca}/\text{Q}_\text{dir}$, tracing the orbital trend of $\text{U}_\text{sca}$ (Fig.~\ref{fig:broadband}c, orange), which is antisymmetric about phase $0.5$.

\subsection{Energy dependence of the polarisation modulation}
\label{sec:energyDependence}
\begin{figure}
    \centering
	\includegraphics[trim={0 0 0 0}, clip,width=\columnwidth]{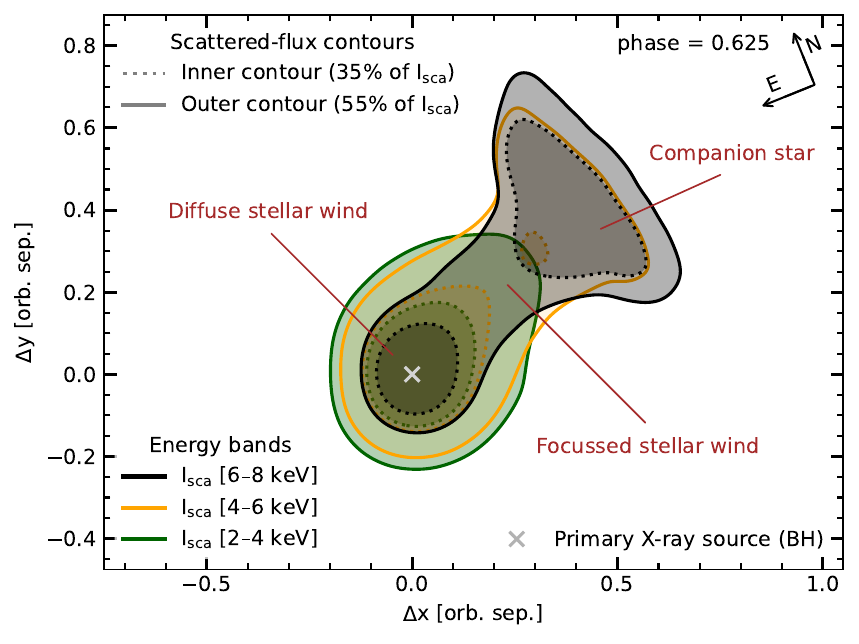}
    \caption{Energy dependence of the extended scattered-flux region: with increasing energy, the dominant scattering region shifts from the diffuse stellar wind close to the X-ray source (green) to the focussed stellar wind and companion star surface (black). The inner and outer contours encircle $35\%$ and $55\%$ of the scattered flux $\text{I}_\text{sca}$, respectively.}
    \label{fig:contoursEnergy}
\end{figure}
The previous subsections focussed on the $6-8~\text{keV}$ band, which exhibits the most prominent polarisation modulation (Fig.~\ref{fig:orbitalModulation}), with a peak-to-peak PD amplitude of $1.24$ percentage points, compared to $0.81$ and $0.25$ in the $4-6$ and $2-4~\text{keV}$ bands, respectively. The modulation strength increases with energy, a trend that continues beyond the IXPE range and reaches $3.80$ percentage points in the {\it XL-Calibur} band ($19-64~\text{keV}$).

This energy dependence reflects the growing influence of the scattered flux on the total observed polarisation signal as energy increases. First, the scattered emission travels longer distances through the binary medium than the direct source emission, and is therefore disproportionately suppressed by photo-absorption; as the absorption efficiency drops steeply across the $2-8~\text{keV}$ band \citep[e.g.][]{vandermeulen23}, the scattered-to-direct flux ratio rises. Since the PD modulation originates entirely from the scattered flux (with $\text{PD}_\text{dir}=\text{PD}_\text{inp}$ constant; Sect.~\ref{sec:RTresults}), the modulation amplitude scales with this flux ratio, and therefore grows with energy.

Second, a similar reasoning holds for the two spatial regions that contribute to the scattered flux: (i) the diffuse scattering halo centred on the X-ray source, which induces low net polarisation, and (ii) the more distant reflection off the companion star and the focussed stellar wind, which drives the prominent double-peaked PD modulation (Sect.~\ref{sec:reflection}). As the latter region corresponds to longer paths through parts of the binary medium that are also denser, it is more strongly attenuated by photo-absorption; its relative contribution therefore rises with energy, reinforcing the PD modulation beyond the rising scattered-to-direct flux ratio. This is visualised in Fig.~\ref{fig:contoursEnergy}: across the IXPE band, the scattered flux progressively shifts from the central scattering halo to the more distant companion star surface with increasing energy.

\subsection{Reduction of the phase-averaged polarisation degree}
\label{sec:reduction}
Finally, we address why the phase-averaged polarisation degree is reduced relative to the intrinsic source polarisation (Sect.~\ref{sec:RTresults}). This reduction amounts to $0.17$, $0.22$, and $0.28$ percentage points in the $2-4$, $4-6$, and $6-8~\text{keV}$ bands, relative to $\text{PD}_\text{inp} = 3.8\%$, $4.4\%$, and $4.8\%$, respectively. In relative terms, the polarisation degree is reduced by a notable $5\%$, $5\%$, and $6\%$.

First, the anisotropic X-ray source (Fig.~\ref{fig:kerrC}) disproportionately illuminates the polar regions of the diffuse stellar wind, inducing a net polarisation signal perpendicular to the source polarisation direction. As the diffuse scattering halo is visible at all orbital phases (Fig.~\ref{fig:images}), this signal persistently counteracts the source polarisation, reducing the phase-averaged PD.

Second, the distant-reflection contribution (driving the main polarisation modulation) peaks when the illuminated face of the companion star is maximally exposed (near phase $0.5$; Fig.~\ref{fig:broadband}e). At this phase, the reflection-induced polarisation signal counteracts the source polarisation (Fig.~\ref{fig:polmap}). Near phases $0.25$ and $0.75$, where the source polarisation is amplified, the polarised flux is much weaker, so the net reduction near phase $0.5$ dominates the phase average. Moreover, the distant reflection contributes a strictly positive Stokes I flux at every orbital phase (Fig.~\ref{fig:broadband}d), diluting the phase-averaged polarisation degree even when the polarised flux modulation partially cancels.

\subsection{Stokes parameters in the IXPE reference frame}
\label{sec:roll}
\begin{figure}[t]
    \centering
	\includegraphics[trim={0 0 0 0}, clip, width=\columnwidth]{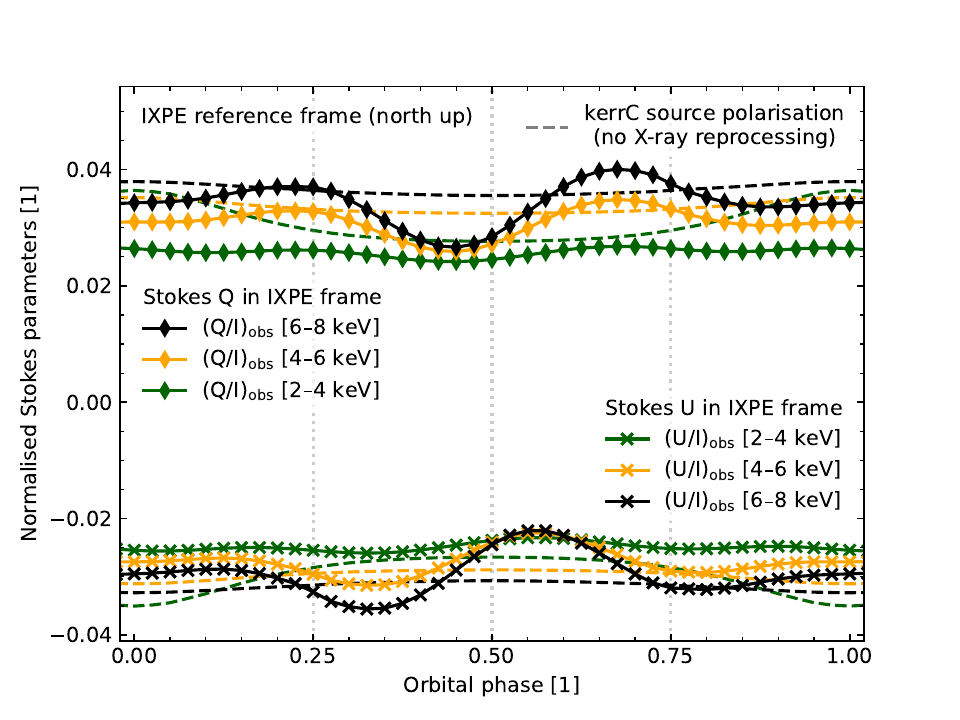}
    \caption{Normalised Stokes parameters $(\text{Q}/\text{I})_\text{obs}$ and $(\text{U}/\text{I})_\text{obs}$ in the IXPE reference frame (celestial north up), accounting for the $21^{\circ}$ roll angle between the projected accretion-disk axis and celestial north (Sect.~\ref{sec:roll}).}
    \label{fig:obsQU}
\end{figure}
Throughout this work, we have oriented the observer such that the sky-projected accretion-disk axis points along the positive $y$-direction. This choice has greatly simplified our analysis: the source polarisation aligns with the positive Stokes Q-direction, and the PD and PA modulations trace $\text{Q}_\text{sca}$ and $\text{U}_\text{sca}$, respectively (Sect.~\ref{sec:QUplane}).

IXPE observations, however, are referred to celestial north. As the projected accretion-disk axis of Cyg X-1 is offset from north by $21^{\circ}$ towards the west \citep{krawczynski22b}, the observed Stokes $\text{Q}_\text{obs}$ and $\text{U}_\text{obs}$ parameters are obtained by rotating $\text{Q}_\text{tot}$ and $\text{U}_\text{tot}$ (Fig.~\ref{fig:broadband}b,c, red) around the line of sight:
\begin{subequations}\label{eq:QUobs}
\begin{align}
\text{Q}_\text{obs} &= \phantom{-}\cos\left(2 \times 21^{\circ}\right) \text{Q}_\text{tot} + \sin\left(2 \times 21^{\circ}\right) \text{U}_\text{tot},\\
\text{U}_\text{obs} &= -\sin\left(2 \times 21^{\circ}\right) \text{Q}_\text{tot} + \cos\left(2 \times 21^{\circ}\right) \text{U}_\text{tot}.
\end{align}
\end{subequations}
The normalised $(\text{Q}/\text{I})_\text{obs}$ and $(\text{U}/\text{I})_\text{obs}$ parameters are shown in Fig.~\ref{fig:obsQU}, which can be directly compared to the Stokes Q and U parameters as measured by IXPE. We note that the main results of this work, the PD and $\Delta\text{PA}$ modulations (Figs.~\ref{fig:orbitalModulation} and~\ref{fig:deltaGamma}), are invariant under this change of reference direction.

\section{Summary and outlook}
\label{sec:summary}
This work presents a 3D radiative transfer model for Cyg \mbox{X-1}, combining a detailed description of the polarised X-ray source emission (Sect.~\ref{sec:kerrC}) with a focussed stellar wind model for the binary medium (Sect.~\ref{sec:focussedwind}). Using the 3D X-ray radiative transfer code SKIRT \citep{vandermeulen23, vandermeulen24b}, we model X-ray reprocessing in the system, and simulate broadband Stokes I, Q, and U fluxes, surface brightness maps, and polarisation maps over one binary orbit, to quantify the reflection-induced PD and PA variations as a function of orbital phase and photon energy.

We find a prominent double-peaked ($P_\mathrm{orb}/2$) polarisation modulation, with a peak-to-peak PD amplitude of $0.25$, $0.81$, and $1.24$ percentage points in the $2-4$, $4-6$, and $6-8~\text{keV}$ bands, respectively, which corresponds to a substantial modulation of the source polarisation degree ($7\%$, $19\%$, and $26\%$; Fig.~\ref{fig:orbitalModulation}). The PA modulation is more modest, with $|\Delta\text{PA}| < 4.6^\circ$ (Fig.~\ref{fig:deltaGamma}). Yet, this PA modulation could constrain the 
sense of the orbital rotation of Cyg X-1, which is currently established only through interferometric radio astrometry \citep[][and references therein]{millerjones21}; an independent confirmation through X-ray polarimetry would be valuable.

The polarisation modulation originates from reflection off the companion star and the focussed stellar wind (Sect.~\ref{sec:reflection}), whose reflection-induced polarisation signal is perpendicular to the donor-black hole axis (Fig.~\ref{fig:polmap}), and rotates with the binary orbit, alternately reinforcing and counteracting the direct source polarisation. This naturally produces the double-peaked $P_\mathrm{orb}/2$ periodicity (Fig.~\ref{fig:orbitalModulation}). The PD modulation amplitude increases with energy (Sect.~\ref{sec:energyDependence}), which is related to the penetrating power of the X-ray photons: as the modulation arises from reflection off the distant star and wind, which corresponds to longer and denser paths through the binary medium, the distant-reflection signal is more prominent at higher energies, where the extinction is lower. 
The bulk of the scattered flux shifts from the central scattering halo (low $\text{PD}$) to the distant companion star surface (high $\text{PD}$) with increasing energy (Fig.~\ref{fig:contoursEnergy}).

Crucially, X-ray reflection also reduces the phase-averaged polarisation degree relative to the intrinsic source polarisation, by $5\%$, $5\%$, and $6\%$ in the three energy bands (Sect.~\ref{sec:reduction}). This is because reflection off the diffuse stellar wind counteracts the source polarisation at all phases, and because the distant-reflection contribution (i.e., the main polarisation modulation) peaks when counteracting the source polarisation (Fig.~\ref{fig:polmap}). We note that such reduction should be accounted for in all wind-fed X-ray binaries, where reflection within the binary system is expected to systematically lower the observed polarisation. For BHXBs, this reduction further increases the tension between the unexpectedly high PDs measured by IXPE and those predicted by theoretical disk-corona models \citep[][]{krawczynski22b}.

This work predicts a polarisation modulation with a period of $P_\mathrm{orb}/2$, induced by reflection off the companion star and its wind. \citet{kravtsov25} recently reported tentative evidence for a PD modulation with the orbital period $P_\mathrm{orb}$, but did not quantify the amplitude of any modulation at $P_\mathrm{orb}/2$. They analysed the full $2-8~\text{keV}$ band, which is dominated by the $2-4~\text{keV}$ flux, where we predict a $P_\mathrm{orb}/2$ modulation amplitude (peak-to-peak) of only 0.25 percentage points (Fig.~\ref{fig:orbitalModulation}), which is well below the uncertainties of the IXPE data. Our predictions therefore do not contradict the observational results of \citet{kravtsov25}.

Moreover, we note that the intrinsic source polarisation of Cyg X-1 varies with accretion state, on timescales comparable to the orbital period \citep[][]{tananbaum72, pottschmidt03, done07, grinberg13}; these intrinsic variations can mask or mimic the reflection-induced PD and PA modulations, complicating a detection. In a forthcoming paper, we confront our predictions with the accumulated IXPE data, by analysing the phase-dependent polarisation measurements in three energy bands and isolating the orbital signal from spectral-state variations, to constrain the reflection-induced modulation and its detectability with IXPE.

The SKIRT simulations presented in this work generated a large, quantitative dataset of phase-resolved Stokes I, Q, and U broadband fluxes, spectra, and surface brightness maps over the $1-100~\text{keV}$ range. This enables a range of follow-up studies, including spatially resolved and spectral analyses, and extensions to higher-energy bands such as {\it XL-Calibur} ($19-64~\text{keV}$; \citealt{abarr21}). Furthermore, it provides a theoretical basis for the interpretation of ongoing IXPE campaigns on BHXBs, and quantitative predictions for current and future X-ray spectro-polarimetric missions.

Our radiative transfer model for Cyg X-1 can be extended in several directions. First, the stellar wind of HDE~226868 is highly structured \citep{Owocki_1984a, grinberg15, Sundqvist_2018a, Hirsch_2019a}. Wind clumping is readily incorporated into SKIRT \citep[see e.g.][]{stalevski16}, but disentangling its polarisation signature warrants a dedicated study, and is deferred to future work. Second, the current model considers only the wind density distribution; the corresponding velocity field can also be imported in SKIRT \citep{camps20}, introducing velocity shifts and broadening that are below the spectral resolution of IXPE, but are directly accessible with XRISM/Resolve spectroscopy \citep{tashiro25}. Finally, a more detailed treatment of the ionisation structure is foreseen with the forthcoming SKIRT release of X-ray radiative transfer in partially ionised gas.

The presented framework can also be applied to Cyg X-1 in
the soft (or intermediate) accretion state, or could be extended to other wind-fed X-ray binaries, such as the neutron-star systems Vela X-1 and Cen X-3. All simulation elements used in this work (such as the interfaces for importing custom sources and transfer media) are built into the public master branch of the SKIRT code, which is publicly available and well documented\footnote{\url{https://skirt.ugent.be}}. We warmly invite the community to use the code in any way they see fit.

\begin{acknowledgements}
       B.\,V.\ acknowledges support through the European Space Agency (ESA) Research Fellowship in Space Science, and thanks Ghent University and SRON for their hospitality throughout the course of this work. K.\,H.\ and H.\,K.\ thank NASA for support under the grants 
80NSSC24K1178, 80NSSC24K1749, and 80NSSC24K1819, and acknowledge support from the McDonnell Center for the Space Sciences at Washington University in St. Louis. They furthermore acknowledge the use of the High Performance Computing cluster of the WashU Physics Department maintained by S.\,Iyer.

\end{acknowledgements}
\bibliographystyle{aa}
\bibliography{finalrefs}

@ARTICLE{abarr21,
       author = {{Abarr}, Q. and {Awaki}, H. and {Baring}, M.~G. and {Bose}, R. and {De Geronimo}, G. and {Dowkontt}, P. and {Errando}, M. and {Guarino}, V. and {Hattori}, K. and {Hayashida}, K. and {Imazato}, F. and {Ishida}, M. and {Iyer}, N.~K. and {Kislat}, F. and {Kiss}, M. and {Kitaguchi}, T. and {Krawczynski}, H. and {Lisalda}, L. and {Matake}, H. and {Maeda}, Y. and {Matsumoto}, H. and {Mineta}, T. and {Miyazawa}, T. and {Mizuno}, T. and {Okajima}, T. and {Pearce}, M. and {Rauch}, B.~F. and {Ryde}, F. and {Shreves}, C. and {Spooner}, S. and {Stana}, T.-A. and {Takahashi}, H. and {Takeo}, M. and {Tamagawa}, T. and {Tamura}, K. and {Tsunemi}, H. and {Uchida}, N. and {Uchida}, Y. and {West}, A.~T. and {Wulf}, E.~A. and {Yamamoto}, R.},
        title = "{XL-Calibur - a second-generation balloon-borne hard X-ray polarimetry mission}",
      journal = {Astroparticle Physics},
         year = 2021,
        month = mar,
       volume = {126},
          eid = {102529},
        pages = {102529},
          doi = {10.1016/j.astropartphys.2020.102529},
archivePrefix = {arXiv},
       eprint = {2010.10608},
 primaryClass = {astro-ph.IM},
       adsurl = {https://ui.adsabs.harvard.edu/abs/2021APh...12602529A}
}

@ARTICLE{niedzwiecki26,
       author = {{Niedzwiecki}, A. and {Szanecki}, M. and {Veledina}, A. and {Zdziarski}, A.~A. and {Chakraborty}, A. and {Poutanen}, J. and {Lubinski}, P. and {Salganik}, A.},
        title = "{X-ray polarization in the soft state of Cyg X-1}",
      journal = {arXiv e-prints},
         year = 2026,
        month = mar,
          eid = {arXiv:2603.10870},
        pages = {arXiv:2603.10870},
          doi = {10.48550/arXiv.2603.10870},
archivePrefix = {arXiv},
       eprint = {2603.10870},
 primaryClass = {astro-ph.HE},
       adsurl = {https://ui.adsabs.harvard.edu/abs/2026arXiv260310870N}
}

@ARTICLE{done07,
       author = {{Done}, Chris and {Gierli{\'n}ski}, Marek and {Kubota}, Aya},
        title = "{Modelling the behaviour of accretion flows in X-ray binaries. Everything you always wanted to know about accretion but were afraid to ask}",
      journal = {\aapr},
         year = 2007,
        month = dec,
       volume = {15},
       number = {1},
        pages = {1-66},
          doi = {10.1007/s00159-007-0006-1},
archivePrefix = {arXiv},
       eprint = {0708.0148},
 primaryClass = {astro-ph},
       adsurl = {https://ui.adsabs.harvard.edu/abs/2007A&ARv..15....1D}
}

@ARTICLE{tananbaum72,
       author = {{Tananbaum}, H. and {Gursky}, H. and {Kellogg}, E. and {Giacconi}, R. and {Jones}, C.},
        title = "{Observation of a Correlated X-Ray Transition in Cygnus X-1}",
      journal = {\apjl},
         year = 1972,
        month = oct,
       volume = {177},
        pages = {L5},
          doi = {10.1086/181042},
       adsurl = {https://ui.adsabs.harvard.edu/abs/1972ApJ...177L...5T}
}

@ARTICLE{pottschmidt03,
       author = {{Pottschmidt}, K. and {Wilms}, J. and {Nowak}, M.~A. and {Pooley}, G.~G. and {Gleissner}, T. and {Heindl}, W.~A. and {Smith}, D.~M. and {Remillard}, R. and {Staubert}, R.},
        title = "{Long term variability of Cygnus X-1.  I. X-ray spectral-temporal correlations in the hard state}",
      journal = {\aap},
         year = 2003,
        month = sep,
       volume = {407},
        pages = {1039-1058},
          doi = {10.1051/0004-6361:20030906},
archivePrefix = {arXiv},
       eprint = {astro-ph/0202258},
 primaryClass = {astro-ph},
       adsurl = {https://ui.adsabs.harvard.edu/abs/2003A&A...407.1039P}
}

@ARTICLE{grinberg13,
       author = {{Grinberg}, V. and {Hell}, N. and {Pottschmidt}, K. and {B{\"o}ck}, M. and {Nowak}, M.~A. and {Rodriguez}, J. and {Bodaghee}, A. and {Cadolle Bel}, M. and {Case}, G.~L. and {Hanke}, M. and {K{\"u}hnel}, M. and {Markoff}, S.~B. and {Pooley}, G.~G. and {Rothschild}, R.~E. and {Tomsick}, J.~A. and {Wilson-Hodge}, C.~A. and {Wilms}, J.},
        title = "{Long term variability of Cygnus X-1. V. State definitions with all sky monitors}",
      journal = {\aap},
         year = 2013,
        month = jun,
       volume = {554},
          eid = {A88},
        pages = {A88},
          doi = {10.1051/0004-6361/201321128},
archivePrefix = {arXiv},
       eprint = {1303.1198},
 primaryClass = {astro-ph.HE},
       adsurl = {https://ui.adsabs.harvard.edu/abs/2013A&A...554A..88G}
}

@ARTICLE{brigitte25,
       author = {{Brigitte}, M. and {Hadrava}, P. and {Kub{\'a}tov{\'a}}, B. and {Cabezas}, M. and {Svoboda}, J. and {{\v{S}}lechta}, M. and {Skarka}, M. and {Alabarta}, K. and {Maryeva}, O. and {Russell}, D.~M. and {Baglio}, M.~C.},
        title = "{Disentangling the stellar atmosphere and the focused wind in different accretion states of Cygnus X-1}",
      journal = {\aap},
         year = 2025,
        month = mar,
       volume = {695},
          eid = {A115},
        pages = {A115},
          doi = {10.1051/0004-6361/202450910},
archivePrefix = {arXiv},
       eprint = {2502.07989},
 primaryClass = {astro-ph.HE},
       adsurl = {https://ui.adsabs.harvard.edu/abs/2025A&A...695A.115B}
}

@ARTICLE{rankin24,
       author = {{Rankin}, John and {Kravtsov}, Vadim and {Muleri}, Fabio and {Poutanen}, Juri and {Marin}, Fr{\'e}d{\'e}ric and {Capitanio}, Fiamma and {Matt}, Giorgio and {Costa}, Enrico and {Marco}, Alessandro Di and {Fabiani}, Sergio and {La Monaca}, Fabio and {Marra}, Lorenzo and {Soffitta}, Paolo},
        title = "{X-Ray Polarimetry as a Tool to Constrain Orbital Parameters in X-Ray Binaries}",
      journal = {\apj},
         year = 2024,
        month = feb,
       volume = {962},
       number = {1},
          eid = {34},
        pages = {34},
          doi = {10.3847/1538-4357/ad1991},
archivePrefix = {arXiv},
       eprint = {2312.16967},
 primaryClass = {astro-ph.HE},
       adsurl = {https://ui.adsabs.harvard.edu/abs/2024ApJ...962...34R}
}

@ARTICLE{kravtsov20,
       author = {{Kravtsov}, Vadim and {Berdyugin}, Andrei V. and {Piirola}, Vilppu and {Kosenkov}, Ilia A. and {Tsygankov}, Sergey S. and {Chernyakova}, Maria and {Malyshev}, Denys and {Sakanoi}, Takeshi and {Kagitani}, Masato and {Berdyugina}, Svetlana V. and {Poutanen}, Juri},
        title = "{Orbital variability of the optical linear polarization of the {\ensuremath{\gamma}}-ray binary LS I +61{\textdegree} 303 and new constraints on the orbital parameters}",
      journal = {\aap},
         year = 2020,
        month = nov,
       volume = {643},
          eid = {A170},
        pages = {A170},
          doi = {10.1051/0004-6361/202038745},
archivePrefix = {arXiv},
       eprint = {2010.00999},
 primaryClass = {astro-ph.SR},
       adsurl = {https://ui.adsabs.harvard.edu/abs/2020A&A...643A.170K}
}

@ARTICLE{ahlberg24,
       author = {{Ahlberg}, Varpu and {Kravtsov}, Vadim and {Poutanen}, Juri},
        title = "{Orbital variability of polarized X-ray radiation reflected from a companion star in X-ray binaries}",
      journal = {\aap},
         year = 2024,
        month = aug,
       volume = {688},
          eid = {A220},
        pages = {A220},
          doi = {10.1051/0004-6361/202450131},
archivePrefix = {arXiv},
       eprint = {2403.17644},
 primaryClass = {astro-ph.HE},
       adsurl = {https://ui.adsabs.harvard.edu/abs/2024A&A...688A.220A}
}

@ARTICLE{matsumoto24,
       author = {{Matsumoto}, Kosei and {Hirashita}, Hiroyuki and {Nagamine}, Kentaro and {van der Giessen}, Stefan and {Romano}, Leonard E.~C. and {Rela{\~n}o}, Monica and {De Looze}, Ilse and {Baes}, Maarten and {Nersesian}, Angelos and {Camps}, Peter and {Hou}, Kuan-chou and {Oku}, Yuri},
        title = "{Observational signatures of the dust size evolution in isolated galaxy simulations}",
      journal = {\aap},
         year = 2024,
        month = sep,
       volume = {689},
          eid = {A79},
        pages = {A79},
          doi = {10.1051/0004-6361/202449454},
archivePrefix = {arXiv},
       eprint = {2402.02659},
 primaryClass = {astro-ph.GA},
       adsurl = {https://ui.adsabs.harvard.edu/abs/2024A&A...689A..79M}
}

@ARTICLE{matsumoto25,
       author = {{Matsumoto}, Kosei and {Sommovigo}, Laura and {Gebek}, Andrea and {Nagamine}, Kentaro and {Nersesian}, Angelos and {Baes}, Maarten and {De Looze}, Ilse and {van der Wel}, Arjen and {Somerville}, Rachel and {Romano}, Leonard E.~C. and {Cochrane}, Rachel K.},
        title = "{Evolution of galaxy attenuation curves driven by evolving dust mass and grain size distributions}",
      journal = {arXiv e-prints},
         year = 2025,
        month = aug,
          eid = {arXiv:2508.21157},
        pages = {arXiv:2508.21157},
          doi = {10.48550/arXiv.2508.21157},
archivePrefix = {arXiv},
       eprint = {2508.21157},
 primaryClass = {astro-ph.GA},
       adsurl = {https://ui.adsabs.harvard.edu/abs/2025arXiv250821157M}
}

@ARTICLE{baes24a,
       author = {{Baes}, Maarten and {Gebek}, Andrea and {Tr{\v{c}}ka}, Ana and {Camps}, Peter and {van der Wel}, Arjen and {Abdurro'uf} and {Andreadis}, Nick and {Tulu}, Sena Bokona and {Emana}, Abdissa Tassama and {Fritz}, Jacopo and {Kelly}, Raymond and {Kova{\v{c}}i{\'c}}, Inja and {La Marca}, Antonio and {Martorano}, Marco and {Mosenkov}, Aleksandr and {Nersesian}, Angelos and {Rodriguez-Gomez}, Vicente and {Tortora}, Crescenzo and {Vander Meulen}, Bert and {Wang}, Lingyu},
        title = "{The TNG50-SKIRT Atlas: Post-processing methodology and first data release}",
      journal = {\aap},
         year = 2024,
        month = mar,
       volume = {683},
          eid = {A181},
        pages = {A181},
          doi = {10.1051/0004-6361/202348418},
archivePrefix = {arXiv},
       eprint = {2401.04224},
 primaryClass = {astro-ph.GA},
       adsurl = {https://ui.adsabs.harvard.edu/abs/2024A&A...683A.181B}
}

@ARTICLE{baes24b,
       author = {{Baes}, Maarten and {Mosenkov}, Aleksandr and {Kelly}, Raymond and {Abdurro'uf} and {Andreadis}, Nick and {Bokona Tulu}, Sena and {Camps}, Peter and {Tassama Emana}, Abdissa and {Fritz}, Jacopo and {Gebek}, Andrea and {Kova{\v{c}}i{\'c}}, Inja and {La Marca}, Antonio and {Martorano}, Marco and {Nersesian}, Angelos and {Rodriguez-Gomez}, Vicente and {Tortora}, Crescenzo and {Tr{\v{c}}ka}, Ana and {Vander Meulen}, Bert and {van der Wel}, Arjen and {Wang}, Lingyu},
        title = "{The TNG50-SKIRT Atlas: Wavelength dependence of the effective radius}",
      journal = {\aap},
         year = 2024,
        month = mar,
       volume = {683},
          eid = {A182},
        pages = {A182},
          doi = {10.1051/0004-6361/202348419},
archivePrefix = {arXiv},
       eprint = {2401.04225},
 primaryClass = {astro-ph.GA},
       adsurl = {https://ui.adsabs.harvard.edu/abs/2024A&A...683A.182B}
}

@ARTICLE{kapoor24,
       author = {{Kapoor}, Anand Utsav and {Baes}, Maarten and {van der Wel}, Arjen and {Gebek}, Andrea and {Camps}, Peter and {Smith}, Aaron and {Boquien}, M{\'e}d{\'e}ric and {Andreadis}, Nick and {Vicens}, Sebastien},
        title = "{TODDLERS: A new UV-millimeter emission library for star-forming regions: II. Star-formation rate indicators using Auriga zoom simulations}",
      journal = {\aap},
         year = 2024,
        month = dec,
       volume = {692},
          eid = {A79},
        pages = {A79},
          doi = {10.1051/0004-6361/202451207},
archivePrefix = {arXiv},
       eprint = {2410.01067},
 primaryClass = {astro-ph.GA},
       adsurl = {https://ui.adsabs.harvard.edu/abs/2024A&A...692A..79K}
}

@ARTICLE{gebek24,
       author = {{Gebek}, Andrea and {Tr{\v{c}}ka}, Ana and {Baes}, Maarten and {Martorano}, Marco and {Pillepich}, Annalisa and {Kapoor}, Anand Utsav and {Nersesian}, Angelos and {van der Wel}, Arjen},
        title = "{The many colours of the TNG100 simulation}",
      journal = {\mnras},
         year = 2024,
        month = jul,
       volume = {531},
       number = {4},
        pages = {3839-3857},
          doi = {10.1093/mnras/stae1377},
archivePrefix = {arXiv},
       eprint = {2405.04925},
 primaryClass = {astro-ph.GA},
       adsurl = {https://ui.adsabs.harvard.edu/abs/2024MNRAS.531.3839G}
}

@ARTICLE{ramachandran25,
       author = {{Ramachandran}, V. and {Sander}, A.~A.~C. and {Oskinova}, L.~M. and {Sch{\"o}sser}, E.~C. and {Pauli}, D. and {Hamann}, W.-R. and {Mahy}, L. and {Bernini-Peron}, M. and {Brigitte}, M. and {Kub{\'a}tov{\'a}}, B.},
        title = "{Comprehensive UV and optical spectral analysis of Cygnus X-1: Stellar and wind parameters, abundances, and evolutionary implications}",
      journal = {\aap},
         year = 2025,
        month = jun,
       volume = {698},
          eid = {A37},
        pages = {A37},
          doi = {10.1051/0004-6361/202554184},
archivePrefix = {arXiv},
       eprint = {2504.05885},
 primaryClass = {astro-ph.SR},
       adsurl = {https://ui.adsabs.harvard.edu/abs/2025A&A...698A..37R}
}

@ARTICLE{west23,
       author = {{West}, Andrew Thomas and {Krawczynski}, Henric},
        title = "{Impact of the Accretion Disk Thickness on the Polarization of the Thermal Emission from Stellar Mass Black Holes}",
      journal = {\apj},
         year = 2023,
        month = nov,
       volume = {957},
       number = {1},
          eid = {9},
        pages = {9},
          doi = {10.3847/1538-4357/acf612},
       adsurl = {https://ui.adsabs.harvard.edu/abs/2023ApJ...957....9W}
}

@ARTICLE{cavero23,
       author = {{Rodriguez Cavero}, Nicole and {Marra}, Lorenzo and {Krawczynski}, Henric and {Dov{\v{c}}iak}, Michal and {Bianchi}, Stefano and {Steiner}, James F. and {Svoboda}, Jiri and {Capitanio}, Fiamma and {Matt}, Giorgio and {Negro}, Michela and {Ingram}, Adam and {Veledina}, Alexandra and {Taverna}, Roberto and {Karas}, Vladimir and {Ursini}, Francesco and {Podgorn{\'y}}, Jakub and {Ratheesh}, Ajay and {Suleimanov}, Valery and {Miku{\v{s}}incov{\'a}}, Romana and {Zane}, Silvia and {Kaaret}, Philip and {Muleri}, Fabio and {Poutanen}, Juri and {Malacaria}, Christian and {Petrucci}, Pierre-Olivier and {Gau}, Ephraim and {Hu}, Kun and {Chun}, Sohee and {Agudo}, Iv{\'a}n and {Antonelli}, Lucio A. and {Bachetti}, Matteo and {Baldini}, Luca and {Baumgartner}, Wayne H. and {Bellazzini}, Ronaldo and {Bongiorno}, Stephen D. and {Bonino}, Raffaella and {Brez}, Alessandro and {Bucciantini}, Niccol{\`o} and {Castellano}, Simone and {Cavazzuti}, Elisabetta and {Chen}, Chien-Ting and {Ciprini}, Stefano and {Costa}, Enrico and {De Rosa}, Alessandra and {Del Monte}, Ettore and {Di Gesu}, Laura and {Di Lalla}, Niccol{\`o} and {Di Marco}, Alessandro and {Donnarumma}, Immacolata and {Doroshenko}, Victor and {Ehlert}, Steven R. and {Enoto}, Teruaki and {Evangelista}, Yuri and {Fabiani}, Sergio and {Ferrazzoli}, Riccardo and {Garc{\'\i}a}, Javier A. and {Gunji}, Shuichi and {Hayashida}, Kiyoshi and {Heyl}, Jeremy and {Iwakiri}, Wataru and {Jorstad}, Svetlana G. and {Kislat}, Fabian and {Kitaguchi}, Takao and {Kolodziejczak}, Jeffery J. and {La Monaca}, Fabio and {Latronico}, Luca and {Liodakis}, Ioannis and {Maldera}, Simone and {Manfreda}, Alberto and {Marin}, Fr{\'e}d{\'e}ric and {Marinucci}, Andrea and {Marscher}, Alan P. and {Marshall}, Herman L. and {Massaro}, Francesco and {Mitsuishi}, Ikuyuki and {Mizuno}, Tsunefumi and {Ng}, Chi-Yung and {O'Dell}, Stephen L. and {Omodei}, Nicola and {Oppedisano}, Chiara and {Papitto}, Alessandro and {Pavlov}, George G. and {Peirson}, Abel L. and {Perri}, Matteo and {Pesce-Rollins}, Melissa and {Pilia}, Maura and {Possenti}, Andrea and {Puccetti}, Simonetta and {Ramsey}, Brian D. and {Rankin}, John and {Roberts}, Oliver J. and {Romani}, Roger W. and {Sgr{\`o}}, Carmelo and {Slane}, Patrick and {Spandre}, Gloria and {Soffitta}, Paolo and {Swartz}, Douglas A. and {Tamagawa}, Toru and {Tavecchio}, Fabrizio and {Tawara}, Yuzuru and {Tennant}, Allyn F. and {Thomas}, Nicholas E. and {Tombesi}, Francesco and {Trois}, Alessio and {Tsygankov}, Sergey S. and {Turolla}, Roberto and {Vink}, Jacco and {Weisskopf}, Martin C. and {Wu}, Kinwah and {Xie}, Fei},
        title = "{The First X-Ray Polarization Observation of the Black Hole X-Ray Binary 4U 1630-47 in the Steep Power-law State}",
      journal = {\apjl},
         year = 2023,
        month = nov,
       volume = {958},
       number = {1},
          eid = {L8},
        pages = {L8},
          doi = {10.3847/2041-8213/acfd2c},
archivePrefix = {arXiv},
       eprint = {2305.10630},
 primaryClass = {astro-ph.HE},
       adsurl = {https://ui.adsabs.harvard.edu/abs/2023ApJ...958L...8R}
}

@ARTICLE{marra24,
       author = {{Marra}, L. and {Brigitte}, M. and {Rodriguez Cavero}, N. and {Chun}, S. and {Steiner}, J.~F. and {Dov{\v{c}}iak}, M. and {Nowak}, M. and {Bianchi}, S. and {Capitanio}, F. and {Ingram}, A. and {Matt}, G. and {Muleri}, F. and {Podgorn{\'y}}, J. and {Poutanen}, J. and {Svoboda}, J. and {Taverna}, R. and {Ursini}, F. and {Veledina}, A. and {De Rosa}, A. and {Garc{\'\i}a}, J.~A. and {Lutovinov}, A.~A. and {Mereminskiy}, I.~A. and {Farinelli}, R. and {Gunji}, S. and {Kaaret}, P. and {Kallman}, T. and {Krawczynski}, H. and {Kan}, Y. and {Hu}, K. and {Marinucci}, A. and {Mastroserio}, G. and {Mikus̆incov{\'a}}, R. and {Parra}, M. and {Petrucci}, P.-O. and {Ratheesh}, A. and {Soffitta}, P. and {Tombesi}, F. and {Zane}, S. and {Agudo}, I. and {Antonelli}, L.~A. and {Bachetti}, M. and {Baldini}, L. and {Baumgartner}, W.~H. and {Bellazzini}, R. and {Bongiorno}, S.~D. and {Bonino}, R. and {Brez}, A. and {Bucciantini}, N. and {Castellano}, S. and {Cavazzuti}, E. and {Chen}, C. and {Ciprini}, S. and {Costa}, E. and {Del Monte}, E. and {Di Gesu}, L. and {Di Lalla}, N. and {Di Marco}, A. and {Donnarumma}, I. and {Doroshenko}, V. and {Ehlert}, S.~R. and {Enoto}, T. and {Evangelista}, Y. and {Fabiani}, S. and {Ferrazzoli}, R. and {Hayashida}, K. and {Heyl}, J. and {Iwakiri}, W. and {Jorstad}, S.~G. and {Karas}, V. and {Kislat}, F. and {Kitaguchi}, T. and {Kolodziejczak}, J.~J. and {La Monaca}, F. and {Latronico}, L. and {Liodakis}, I. and {Maldera}, S. and {Manfreda}, A. and {Marin}, F. and {Marscher}, A.~P. and {Marshall}, H.~L. and {Massaro}, F. and {Mitsuishi}, I. and {Mizuno}, T. and {Negro}, M. and {Ng}, C.~Y. and {O'Dell}, S.~L. and {Omodei}, N. and {Oppedisano}, C. and {Papitto}, A. and {Pavlov}, G.~G. and {Peirson}, A.~L. and {Perri}, M. and {Pesce-Rollins}, M. and {Pilia}, M. and {Possenti}, A. and {Puccetti}, S. and {Ramsey}, B.~D. and {Rankin}, J. and {Roberts}, O.~J. and {Romani}, R.~W. and {Sgr{\`o}}, C. and {Slane}, P. and {Spandre}, G. and {Swartz}, D.~A. and {Tamagawa}, T. and {Tavecchio}, F. and {Tawara}, Y. and {Tennant}, A.~F. and {Thomas}, N.~E. and {Trois}, A. and {Tsygankov}, S.~S. and {Turolla}, R. and {Vink}, J. and {Weisskopf}, M.~C. and {Wu}, K. and {Xie}, F.},
        title = "{IXPE observation confirms a high spin in the accreting black hole 4U 1957+115}",
      journal = {\aap},
         year = 2024,
        month = apr,
       volume = {684},
          eid = {A95},
        pages = {A95},
          doi = {10.1051/0004-6361/202348277},
archivePrefix = {arXiv},
       eprint = {2310.11125},
 primaryClass = {astro-ph.HE},
       adsurl = {https://ui.adsabs.harvard.edu/abs/2024A&A...684A..95M}
}

@ARTICLE{steiner24,
       author = {{Steiner}, James F. and {Nathan}, Edward and {Hu}, Kun and {Krawczynski}, Henric and {Dov{\v{c}}iak}, Michal and {Veledina}, Alexandra and {Muleri}, Fabio and {Svoboda}, Jiri and {Alabarta}, Kevin and {Parra}, Maxime and {Bhargava}, Yash and {Matt}, Giorgio and {Poutanen}, Juri and {Petrucci}, Pierre-Olivier and {Tennant}, Allyn F. and {Baglio}, M. Cristina and {Baldini}, Luca and {Barnier}, Samuel and {Bhattacharyya}, Sudip and {Bianchi}, Stefano and {Brigitte}, Maimouna and {Cabezas}, Mauricio and {Cangemi}, Floriane and {Capitanio}, Fiamma and {Casey}, Jacob and {Rodriguez Cavero}, Nicole and {Castellano}, Simone and {Cavazzuti}, Elisabetta and {Chun}, Sohee and {Churazov}, Eugene and {Costa}, Enrico and {Di Lalla}, Niccol{\`o} and {Di Marco}, Alessandro and {Egron}, Elise and {Ewing}, Melissa and {Fabiani}, Sergio and {Garc{\'\i}a}, Javier A. and {Green}, David A. and {Grinberg}, Victoria and {Hadrava}, Petr and {Ingram}, Adam and {Kaaret}, Philip and {Kislat}, Fabian and {Kitaguchi}, Takao and {Kravtsov}, Vadim and {Kub{\'a}tov{\'a}}, Brankica and {La Monaca}, Fabio and {Latronico}, Luca and {Loktev}, Vladislav and {Malacaria}, Christian and {Marin}, Fr{\'e}d{\'e}ric and {Marinucci}, Andrea and {Maryeva}, Olga and {Mastroserio}, Guglielmo and {Mizuno}, Tsunefumi and {Negro}, Michela and {Omodei}, Nicola and {Podgorn{\'y}}, Jakub and {Rankin}, John and {Ratheesh}, Ajay and {Rhodes}, Lauren and {Russell}, David M. and {{\v{S}}lechta}, Miroslav and {Soffitta}, Paolo and {Spooner}, Sean and {Suleimanov}, Valery and {Tombesi}, Francesco and {Trushkin}, Sergei A. and {Weisskopf}, Martin C. and {Zane}, Silvia and {Zdziarski}, Andrzej A. and {Zhang}, Sixuan and {Zhang}, Wenda and {Zhou}, Menglei and {Agudo}, Iv{\'a}n and {Antonelli}, Lucio A. and {Bachetti}, Matteo and {Baumgartner}, Wayne H. and {Bellazzini}, Ronaldo and {Bongiorno}, Stephen D. and {Bonino}, Raffaella and {Brez}, Alessandro and {Bucciantini}, Niccol{\`o} and {Chen}, Chien-Ting and {Ciprini}, Stefano and {De Rosa}, Alessandra and {Del Monte}, Ettore and {Di Gesu}, Laura and {Donnarumma}, Immacolata and {Doroshenko}, Victor and {Ehlert}, Steven R. and {Enoto}, Teruaki and {Evangelista}, Yuri and {Ferrazzoli}, Riccardo and {Gunji}, Shuichi and {Hayashida}, Kiyoshi and {Heyl}, Jeremy and {Iwakiri}, Wataru and {Jorstad}, Svetlana G. and {Karas}, Vladimir and {Kolodziejczak}, Jeffery J. and {Liodakis}, Ioannis and {Maldera}, Simone and {Manfreda}, Alberto and {Marscher}, Alan P. and {Marshall}, Herman L. and {Massaro}, Francesco and {Mitsuishi}, Ikuyuki and {Ng}, Chi-Yung and {O'Dell}, Stephen L. and {Oppedisano}, Chiara and {Papitto}, Alessandro and {Pavlov}, George G. and {Peirson}, Abel L. and {Perri}, Matteo and {Pesce-Rollins}, Melissa and {Pilia}, Maura and {Possenti}, Andrea and {Puccetti}, Simonetta and {Ramsey}, Brian D. and {Roberts}, Oliver J. and {Romani}, Roger W. and {Sgr{\`o}}, Carmelo and {Slane}, Patrick and {Spandre}, Gloria and {Swartz}, Douglas A. and {Tamagawa}, Toru and {Tavecchio}, Fabrizio and {Taverna}, Roberto and {Tawara}, Yuzuru and {Thomas}, Nicholas E. and {Trois}, Alessio and {Tsygankov}, Sergey S. and {Turolla}, Roberto and {Vink}, Jacco and {Wu}, Kinwah and {Xie}, Fei},
        title = "{An IXPE-led X-Ray Spectropolarimetric Campaign on the Soft State of Cygnus X-1: X-Ray Polarimetric Evidence for Strong Gravitational Lensing}",
      journal = {\apjl},
         year = 2024,
        month = jul,
       volume = {969},
       number = {2},
          eid = {L30},
        pages = {L30},
          doi = {10.3847/2041-8213/ad58e4},
archivePrefix = {arXiv},
       eprint = {2406.12014},
 primaryClass = {astro-ph.HE},
       adsurl = {https://ui.adsabs.harvard.edu/abs/2024ApJ...969L..30S}
}

@ARTICLE{krawczynski24,
       author = {{Krawczynski}, Henric and {Yuan}, Yajie and {Chen}, Alexander Y. and {Hu}, Kun and {Rodriguez Cavero}, Nicole and {Chun}, Sohee and {Gau}, Ephraim and {Steiner}, James F. and {Dov{\v{c}}iak}, Michal},
        title = "{Evaluation of Several Explanations of the Strong X-Ray Polarization of the Black Hole X-Ray Binary 4U 1630-47}",
      journal = {\apjl},
         year = 2024,
        month = dec,
       volume = {977},
       number = {1},
          eid = {L10},
        pages = {L10},
          doi = {10.3847/2041-8213/ad855c},
archivePrefix = {arXiv},
       eprint = {2307.13141},
 primaryClass = {astro-ph.HE},
       adsurl = {https://ui.adsabs.harvard.edu/abs/2024ApJ...977L..10K}
}

@ARTICLE{draghis25,
       author = {{Draghis}, Paul A. and {Miller}, Jon M. and {Kara}, Erin and {Costantini}, Elisa and {Adegoke}, Oluwashina and {Garcia}, Javier A.},
        title = "{A XRISM View of Relativistic Reflection in Cygnus X-1}",
      journal = {arXiv e-prints},
         year = 2025,
        month = nov,
          eid = {arXiv:2511.17338},
        pages = {arXiv:2511.17338},
archivePrefix = {arXiv},
       eprint = {2511.17338},
 primaryClass = {astro-ph.HE},
       adsurl = {https://ui.adsabs.harvard.edu/abs/2025arXiv251117338D}
}

@ARTICLE{beloborodov98,
       author = {{Beloborodov}, Andrei M.},
        title = "{Polarization Change Due to Fast Winds from Accretion Disks}",
      journal = {\apjl},
         year = 1998,
        month = apr,
       volume = {496},
       number = {2},
        pages = {L105-L108},
          doi = {10.1086/311260},
archivePrefix = {arXiv},
       eprint = {astro-ph/9802128},
 primaryClass = {astro-ph},
       adsurl = {https://ui.adsabs.harvard.edu/abs/1998ApJ...496L.105B}
}

@ARTICLE{tomaru24,
       author = {{Tomaru}, Ryota and {Done}, Chris and {Odaka}, Hirokazu},
        title = "{X-ray polarization properties of thermal-radiative disc winds in binary systems}",
      journal = {\mnras},
         year = 2024,
        month = jan,
       volume = {527},
       number = {3},
        pages = {7047-7054},
          doi = {10.1093/mnras/stad3649},
archivePrefix = {arXiv},
       eprint = {2308.07237},
 primaryClass = {astro-ph.HE},
       adsurl = {https://ui.adsabs.harvard.edu/abs/2024MNRAS.527.7047T}
}

@ARTICLE{poutanen23,
       author = {{Poutanen}, Juri and {Veledina}, Alexandra and {Beloborodov}, Andrei M.},
        title = "{Polarized X-Rays from Windy Accretion in Cygnus X-1}",
      journal = {\apjl},
         year = 2023,
        month = may,
       volume = {949},
       number = {1},
          eid = {L10},
        pages = {L10},
          doi = {10.3847/2041-8213/acd33e},
archivePrefix = {arXiv},
       eprint = {2302.11674},
 primaryClass = {astro-ph.HE},
       adsurl = {https://ui.adsabs.harvard.edu/abs/2023ApJ...949L..10P}
}

@ARTICLE{awaki25,
       author = {{Awaki}, Hisamitsu and {Baring}, Matthew G. and {Bose}, Richard and {Casey}, Jacob and {Chun}, Sohee and {Dasgupta}, Adrika and {Galchenko}, Pavel and {Gau}, Ephraim and {Goya}, Kazuho and {Hakamata}, Tomohiro and {Hayashi}, Takayuki and {Heatwole}, Scott and {Hu}, Kun and {Ishi}, Daiki and {Ishida}, Manabu and {Kislat}, Fabian and {Kiss}, M{\'o}zsi and {Klepper}, Kassi and {Krawczynski}, Henric and {Kuramoto}, Haruki and {Lisalda}, Lindsey and {Maeda}, Yoshitomo and {Matsumoto}, Hironori and {Menon}, Shravan Vengalil and {Miyamoto}, Aiko and {Miyamoto}, Asca and {Murakami}, Kaito and {Okajima}, Takashi and {Pearce}, Mark and {Rauch}, Brian and {Rodriguez Cavero}, Nicole and {Shirahama}, Kentaro and {Spooner}, Sean and {Takahashi}, Hiromitsu and {Tamura}, Keisuke and {Uchida}, Yuusuke and {Wimalasena}, Kasun and {Yokota}, Masato and {Yoshimoto}, Marina},
        title = "{XL-Calibur Polarimetry of Cyg X-1 Further Constrains the Origin of Its Hard-state X-Ray Emission}",
      journal = {\apj},
         year = 2025,
        month = nov,
       volume = {994},
       number = {1},
          eid = {37},
        pages = {37},
          doi = {10.3847/1538-4357/ae0f1d},
       adsurl = {https://ui.adsabs.harvard.edu/abs/2025ApJ...994...37A}
}

@ARTICLE{kravtsov25,
       author = {{Kravtsov}, Vadim and {Bocharova}, Anastasiia and {Veledina}, Alexandra and {Poutanen}, Juri and {Hughes}, Andrew K. and {Dov{\v{c}}iak}, Michal and {Egron}, Elise and {Muleri}, Fabio and {Podgorny}, Jakub and {Svoboda}, Ji{\v{r}}i and {Forsblom}, Sofia V. and {Berdyugin}, Andrei V. and {Blinov}, Dmitry and {Bright}, Joe S. and {Carotenuto}, Francesco and {Green}, David A. and {Ingram}, Adam and {Liodakis}, Ioannis and {Mandarakas}, Nikos and {Nitindala}, Anagha P. and {Rhodes}, Lauren and {Trushkin}, Sergei A. and {Tsygankov}, Sergey S. and {Brigitte}, Ma{\"\i}mouna and {Di Marco}, Alessandro and {Iacolina}, Noemi and {Krawczynski}, Henric and {La Monaca}, Fabio and {Loktev}, Vladislav and {Mastroserio}, Guglielmo and {Petrucci}, Pierre-Olivier and {Pilia}, Maura and {Tombesi}, Francesco and {Zdziarski}, Andrzej A.},
        title = "{Variability of X-ray polarization of Cyg X-1}",
      journal = {\aap},
         year = 2025,
        month = sep,
       volume = {701},
          eid = {A115},
        pages = {A115},
          doi = {10.1051/0004-6361/202555411},
archivePrefix = {arXiv},
       eprint = {2505.03942},
 primaryClass = {astro-ph.HE},
       adsurl = {https://ui.adsabs.harvard.edu/abs/2025A&A...701A.115K}
}

@ARTICLE{miskovicova2016,
       author = {{Mi{\v{s}}kovi{\v{c}}ov{\'a}}, Ivica and {Hell}, Natalie and {Hanke}, Manfred and {Nowak}, Michael A. and {Pottschmidt}, Katja and {Schulz}, Norbert S. and {Grinberg}, Victoria and {Duro}, Refiz and {Madej}, Oliwia K. and {Lohfink}, Anne M. and {Rodriguez}, J{\'e}r{\^o}me and {Cadolle Bel}, Marion and {Bodaghee}, Arash and {Tomsick}, John A. and {Lee}, Julia C. and {Brown}, Gregory V. and {Wilms}, J{\"o}rn},
        title = "{Chandra X-ray spectroscopy of focused wind in the Cygnus X-1 system. II. The non-dip spectrum in the low/hard state - modulations with orbital phase}",
      journal = {\aap},
         year = 2016,
        month = may,
       volume = {590},
          eid = {A114},
        pages = {A114},
          doi = {10.1051/0004-6361/201322490},
archivePrefix = {arXiv},
       eprint = {1604.00364},
 primaryClass = {astro-ph.HE},
       adsurl = {https://ui.adsabs.harvard.edu/abs/2016A&A...590A.114M}
}

@ARTICLE{castor75,
       author = {{Castor}, J.~I. and {Abbott}, D.~C. and {Klein}, R.~I.},
        title = "{Radiation-driven winds in Of stars.}",
      journal = {\apj},
         year = 1975,
        month = jan,
       volume = {195},
        pages = {157-174},
          doi = {10.1086/153315},
       adsurl = {https://ui.adsabs.harvard.edu/abs/1975ApJ...195..157C}
}

@ARTICLE{krawczynski22b,
       author = {{Krawczynski}, Henric and {Muleri}, Fabio and {Dov{\v{c}}iak}, Michal and {Veledina}, Alexandra and {Rodriguez Cavero}, Nicole and {Svoboda}, Jiri and {Ingram}, Adam and {Matt}, Giorgio and {Garcia}, Javier A. and {Loktev}, Vladislav and {Negro}, Michela and {Poutanen}, Juri and {Kitaguchi}, Takao and {Podgorn{\'y}}, Jakub and {Rankin}, John and {Zhang}, Wenda and {Berdyugin}, Andrei and {Berdyugina}, Svetlana V. and {Bianchi}, Stefano and {Blinov}, Dmitry and {Capitanio}, Fiamma and {Di Lalla}, Niccol{\`o} and {Draghis}, Paul and {Fabiani}, Sergio and {Kagitani}, Masato and {Kravtsov}, Vadim and {Kiehlmann}, Sebastian and {Latronico}, Luca and {Lutovinov}, Alexander A. and {Mandarakas}, Nikos and {Marin}, Fr{\'e}d{\'e}ric and {Marinucci}, Andrea and {Miller}, Jon M. and {Mizuno}, Tsunefumi and {Molkov}, Sergey V. and {Omodei}, Nicola and {Petrucci}, Pierre-Olivier and {Ratheesh}, Ajay and {Sakanoi}, Takeshi and {Semena}, Andrei N. and {Skalidis}, Raphael and {Soffitta}, Paolo and {Tennant}, Allyn F. and {Thalhammer}, Phillipp and {Tombesi}, Francesco and {Weisskopf}, Martin C. and {Wilms}, Joern and {Zhang}, Sixuan and {Agudo}, Iv{\'a}n and {Antonelli}, Lucio A. and {Bachetti}, Matteo and {Baldini}, Luca and {Baumgartner}, Wayne H. and {Bellazzini}, Ronaldo and {Bongiorno}, Stephen D. and {Bonino}, Raffaella and {Brez}, Alessandro and {Bucciantini}, Niccol{\`o} and {Castellano}, Simone and {Cavazzuti}, Elisabetta and {Ciprini}, Stefano and {Costa}, Enrico and {De Rosa}, Alessandra and {Del Monte}, Ettore and {Di Gesu}, Laura and {Di Marco}, Alessandro and {Donnarumma}, Immacolata and {Doroshenko}, Victor and {Ehlert}, Steven R. and {Enoto}, Teruaki and {Evangelista}, Yuri and {Ferrazzoli}, Riccardo and {Gunji}, Shuichi and {Hayashida}, Kiyoshi and {Heyl}, Jeremy and {Iwakiri}, Wataru and {Jorstad}, Svetlana G. and {Karas}, Vladimir and {Kolodziejczak}, Jeffery J. and {La Monaca}, Fabio and {Liodakis}, Ioannis and {Maldera}, Simone and {Manfreda}, Alberto and {Marscher}, Alan P. and {Marshall}, Herman L. and {Mitsuishi}, Ikuyuki and {Ng}, Chi-Yung and {O{\textquoteright}Dell}, Stephen L. and {Oppedisano}, Chiara and {Papitto}, Alessandro and {Pavlov}, George G. and {Peirson}, Abel L. and {Perri}, Matteo and {Pesce-Rollins}, Melissa and {Pilia}, Maura and {Possenti}, Andrea and {Puccetti}, Simonetta and {Ramsey}, Brian D. and {Romani}, Roger W. and {Sgr{\`o}}, Carmelo and {Slane}, Patrick and {Spandre}, Gloria and {Tamagawa}, Toru and {Tavecchio}, Fabrizio and {Taverna}, Roberto and {Tawara}, Yuzuru and {Thomas}, Nicholas E. and {Trois}, Alessio and {Tsygankov}, Sergey and {Turolla}, Roberto and {Vink}, Jacco and {Wu}, Kinwah and {Xie}, Fei and {Zane}, Silvia},
        title = "{Polarized x-rays constrain the disk-jet geometry in the black hole x-ray binary Cygnus X-1}",
      journal = {Science},
         year = 2022,
        month = nov,
       volume = {378},
       number = {6620},
        pages = {650-654},
          doi = {10.1126/science.add5399},
archivePrefix = {arXiv},
       eprint = {2206.09972},
 primaryClass = {astro-ph.HE},
       adsurl = {https://ui.adsabs.harvard.edu/abs/2022Sci...378..650K}
}

@ARTICLE{majumder26,
       author = {{Majumder}, Seshadri and {Kushwaha}, Ankur and {Singh}, Swapnil and {Jayasurya}, Kiran M. and {Das}, Santabrata and {Nandi}, Anuj},
        title = "{Probing the accretion geometry of black hole X-ray binaries: a multimission spectro-polarimetric and timing study}",
      journal = {\mnras},
         year = 2026,
        month = jan,
       volume = {545},
       number = {2},
          eid = {staf1933},
        pages = {staf1933},
          doi = {10.1093/mnras/staf1933},
archivePrefix = {arXiv},
       eprint = {2506.03774},
 primaryClass = {astro-ph.HE},
       adsurl = {https://ui.adsabs.harvard.edu/abs/2026MNRAS.545f1933M}
}

@ARTICLE{Dovciak24,
       author = {{Dov{\v{c}}iak}, Michal and {Podgorn{\'y}}, Jakub and {Svoboda}, Ji{\v{r}}{\'\i} and {Steiner}, James F. and {Kaaret}, Philip and {Krawczynski}, Henric and {Ingram}, Adam and {Kravtsov}, Vadim and {Marra}, Lorenzo and {Muleri}, Fabio and {Garc{\'\i}a}, Javier A. and {Mastroserio}, Guglielmo and {Miku{\v{s}}incov{\'a}}, Romana and {Ratheesh}, Ajay and {Cavero}, Nicole Rodriguez},
        title = "{IXPE View of BH XRBs during the First 2.5 Years of the Mission}",
      journal = {Galaxies},
         year = 2024,
        month = sep,
       volume = {12},
       number = {5},
          eid = {54},
        pages = {54},
          doi = {10.3390/galaxies12050054},
       adsurl = {https://ui.adsabs.harvard.edu/abs/2024Galax..12...54D}
}

@ARTICLE{2008ApJ...678.1237G,
       author = {{Gies}, D.~R. and {Bolton}, C.~T. and {Blake}, R.~M. and {Caballero-Nieves}, S.~M. and {Crenshaw}, D.~M. and {Hadrava}, P. and {Herrero}, A. and {Hillwig}, T.~C. and {Howell}, S.~B. and {Huang}, W. and {Kaper}, L. and {Koubsk{\'y}}, P. and {McSwain}, M.~V.},
        title = "{Stellar Wind Variations during the X-Ray High and Low States of Cygnus X-1}",
      journal = {\apj},
         year = 2008,
        month = may,
       volume = {678},
       number = {2},
        pages = {1237-1247},
          doi = {10.1086/586690},
archivePrefix = {arXiv},
       eprint = {0801.4286},
 primaryClass = {astro-ph},
       adsurl = {https://ui.adsabs.harvard.edu/abs/2008ApJ...678.1237G}
}

@ARTICLE{millerjones21,
       author = {{Miller-Jones}, James C.~A. and {Bahramian}, Arash and {Orosz}, Jerome A. and {Mandel}, Ilya and {Gou}, Lijun and {Maccarone}, Thomas J. and {Neijssel}, Coenraad J. and {Zhao}, Xueshan and {Zi{\'o}{\l}kowski}, Janusz and {Reid}, Mark J. and {Uttley}, Phil and {Zheng}, Xueying and {Byun}, Do-Young and {Dodson}, Richard and {Grinberg}, Victoria and {Jung}, Taehyun and {Kim}, Jeong-Sook and {Marcote}, Benito and {Markoff}, Sera and {Rioja}, Mar{\'\i}a J. and {Rushton}, Anthony P. and {Russell}, David M. and {Sivakoff}, Gregory R. and {Tetarenko}, Alexandra J. and {Tudose}, Valeriu and {Wilms}, Joern},
        title = "{Cygnus X-1 contains a 21-solar mass black hole{\textemdash}Implications for massive star winds}",
      journal = {Science},
         year = 2021,
        month = mar,
       volume = {371},
       number = {6533},
        pages = {1046-1049},
          doi = {10.1126/science.abb3363},
archivePrefix = {arXiv},
       eprint = {2102.09091},
 primaryClass = {astro-ph.HE},
       adsurl = {https://ui.adsabs.harvard.edu/abs/2021Sci...371.1046M}
}

@ARTICLE{krawczynski25,
       author = {{Krawczynski}, Henric and {Hu}, Kun},
        title = "{The Cygnus X-1 Puzzle: Implications of X-Ray Polarization Measurements in the Soft and Hard States on the Properties of the Accretion Flow and the Emission Mechanisms}",
      journal = {\apj},
         year = 2025,
        month = nov,
       volume = {993},
       number = {1},
          eid = {54},
        pages = {54},
          doi = {10.3847/1538-4357/adfddf},
archivePrefix = {arXiv},
       eprint = {2506.01184},
 primaryClass = {astro-ph.HE},
       adsurl = {https://ui.adsabs.harvard.edu/abs/2025ApJ...993...54K}
}

@ARTICLE{krawczynski22a,
       author = {{Krawczynski}, H. and {Beheshtipour}, B.},
        title = "{New Constraints on the Spin of the Black Hole Cygnus X-1 and the Physical Properties of its Accretion Disk Corona}",
      journal = {\apj},
         year = 2022,
        month = jul,
       volume = {934},
       number = {1},
          eid = {4},
        pages = {4},
          doi = {10.3847/1538-4357/ac7725},
archivePrefix = {arXiv},
       eprint = {2201.07360},
 primaryClass = {astro-ph.HE},
       adsurl = {https://ui.adsabs.harvard.edu/abs/2022ApJ...934....4K}
}

@ARTICLE{gies86a,
       author = {{Gies}, D.~R. and {Bolton}, C.~T.},
        title = "{The Optical Spectrum of HDE 226868=Cygnus X-1. II. Spectrophotometry and Mass Estimates}",
      journal = {\apj},
         year = 1986,
        month = may,
       volume = {304},
        pages = {371},
          doi = {10.1086/164171},
       adsurl = {https://ui.adsabs.harvard.edu/abs/1986ApJ...304..371G}
}

@ARTICLE{lai24,
       author = {{Lai}, E.~V. and {De Marco}, B. and {Cavecchi}, Y. and {El Mellah}, I. and {Cinus}, M. and {Diez}, C.~M. and {Grinberg}, V. and {Zdziarski}, A.~A. and {Uttley}, P. and {Bachetti}, M. and {Jos{\'e}}, J. and {Sala}, G. and {R{\'o}{\.z}a{\'n}ska}, A. and {Wilms}, J.},
        title = "{Characterisation of the stellar wind in Cyg X-1 via modelling of colour-colour diagrams}",
      journal = {\aap},
         year = 2024,
        month = nov,
       volume = {691},
          eid = {A78},
        pages = {A78},
          doi = {10.1051/0004-6361/202451043},
archivePrefix = {arXiv},
       eprint = {2408.05852},
 primaryClass = {astro-ph.HE},
       adsurl = {https://ui.adsabs.harvard.edu/abs/2024A&A...691A..78L}
}

@ARTICLE{gies86b,
       author = {{Gies}, D.~R. and {Bolton}, C.~T.},
        title = "{The Optical Spectrum of HDE 226868 = Cygnus X-1. III. A Focused Stellar Wind Model for He II lambda 4686 Emission}",
      journal = {\apj},
         year = 1986,
        month = may,
       volume = {304},
        pages = {389},
          doi = {10.1086/164172},
       adsurl = {https://ui.adsabs.harvard.edu/abs/1986ApJ...304..389G}
}

@ARTICLE{brown78,
       author = {{Brown}, J.~C. and {McLean}, I.~S. and {Emslie}, A.~G.},
        title = "{Polarisation by Thomson scattering in optically thin stellar envelopes. II. Binary and multiple star envelopes and the determination of binary inclinations.}",
      journal = {\aap},
         year = 1978,
        month = aug,
       volume = {68},
        pages = {415-427},
       adsurl = {https://ui.adsabs.harvard.edu/abs/1978A&A....68..415B}
}

@ARTICLE{friend82,
       author = {{Friend}, D.~B. and {Castor}, J.~I.},
        title = "{Radiation driven winds in X-ray binaries.}",
      journal = {\apj},
         year = 1982,
        month = oct,
       volume = {261},
        pages = {293-300},
          doi = {10.1086/160340},
       adsurl = {https://ui.adsabs.harvard.edu/abs/1982ApJ...261..293F}
}

@ARTICLE{grinberg15,
       author = {{Grinberg}, V. and {Leutenegger}, M.~A. and {Hell}, N. and {Pottschmidt}, K. and {B{\"o}ck}, M. and {Garc{\'\i}a}, J.~A. and {Hanke}, M. and {Nowak}, M.~A. and {Sundqvist}, J.~O. and {Townsend}, R.~H.~D. and {Wilms}, J.},
        title = "{Long term variability of Cygnus X-1. VII. Orbital variability of the focussed wind in Cyg X-1/HDE 226868 system}",
      journal = {\aap},
         year = 2015,
        month = apr,
       volume = {576},
          eid = {A117},
        pages = {A117},
          doi = {10.1051/0004-6361/201425418},
archivePrefix = {arXiv},
       eprint = {1502.07343},
 primaryClass = {astro-ph.HE},
       adsurl = {https://ui.adsabs.harvard.edu/abs/2015A&A...576A.117G}
}

@ARTICLE{gebek25,
       author = {{Gebek}, Andrea and {Diemer}, Benedikt and {Martorano}, Marco and {van der Wel}, Arjen and {Pantoni}, Lara and {Baes}, Maarten and {Gabrielpillai}, Austen and {Utsav Kapoor}, Anand and {Osinga}, Calvin and {Nersesian}, Angelos and {Matsumoto}, Kosei and {Gordon}, Karl},
        title = "{The mass-dependent UVJ diagram at cosmic noon: A challenge for galaxy evolution models and dust radiative transfer}",
      journal = {\aap},
         year = 2025,
        month = mar,
       volume = {695},
          eid = {A90},
        pages = {A90},
          doi = {10.1051/0004-6361/202452768},
archivePrefix = {arXiv},
       eprint = {2501.12008},
 primaryClass = {astro-ph.GA},
       adsurl = {https://ui.adsabs.harvard.edu/abs/2025A&A...695A..90G}
}

@ARTICLE{baes25,
       author = {{Baes}, M. and {Gebek}, A. and {Kunene}, S. and {Leeuw}, L. and {Nelson}, D. and {Ponomareva}, A.~A. and {Andreadis}, N. and {Bianchetti}, A. and {de Blok}, W.~J.~G. and {Rajohnson}, S.~H.~A. and {Sorgho}, A.},
        title = "{The multi-wavelength Tully-Fisher relation in the TNG50 cosmological simulation}",
      journal = {\aap},
         year = 2025,
        month = apr,
       volume = {696},
          eid = {A52},
        pages = {A52},
          doi = {10.1051/0004-6361/202453417},
archivePrefix = {arXiv},
       eprint = {2503.00194},
 primaryClass = {astro-ph.GA},
       adsurl = {https://ui.adsabs.harvard.edu/abs/2025A&A...696A..52B}
}

@ARTICLE{tashiro25,
       author = {{Tashiro}, Makoto and {Kelley}, Richard and {Watanabe}, Shin and {Maejima}, Hironori and {Reichenthal}, Lillian and {Toda}, Kenichi and {Hartz}, Leslie and {Santovincenzo}, Andrea and {Matsushita}, Kyoko and {Yamaguchi}, Hiroya and {Petre}, Robert and {Williams}, Brian and {Guainazzi}, Matteo and {Costantini}, Elisa and {Takei}, Yoh and {Ishisaki}, Yoshitaka and {Fujimoto}, Ryuichi and {Henegar-Leon}, Joy and {Sneiderman}, Gary and {Tomida}, Hiroshi and {Mori}, Koji and {Nakajima}, Hiroshi and {Terada}, Yukikatsu and {Holland}, Matthew and {Loewenstein}, Michael and {Miller}, Eric and {Sawada}, Makoto and {Kallman}, Timothy and {Kaastra}, Jelle and {Done}, Chris and {Enoto}, Teruaki and {Bamba}, Aya and {Corrales}, Lia and {Ueda}, Yoshihiro and {Kara}, Erin and {Zhuravleva}, Irina and {Fujita}, Yutaka and {Arai}, Yoshitaka and {Audard}, Marc and {Awaki}, Hisamitsu and {Ballhausen}, Ralf and {Baluta}, Chris and {Bando}, Nobutaka and {Behar}, Ehud and {Bialas}, Thomas and {Boissay-Malaquin}, Rozenn and {Brenneman}, Laura and {Brown}, Gregory V. and {Chiao}, Meng and {Cumbee}, Renata and {de Vries}, Cor and {den Herder}, Jan-Willem and {D{\'\i}az Trigo}, Mar{\'\i}a and {DiPirro}, Michael and {Dotani}, Tadayasu and {Carrero}, Jacobo Ebrero and {Ebisawa}, Ken and {Eckart}, Megan and {Eckert}, Dominique and {Eguchi}, Satoshi and {Ezoe}, Yuichiro and {Ferrigno}, Carlo and {Foster}, Adam and {Fukazawa}, Yasushi and {Fukushima}, Kotaro and {Furuzawa}, Akihiro and {Gallo}, Luigi and {Garcia Martinez}, Javier and {Gorter}, Nathalie and {Grim}, Martin and {Gu}, Liyi and {Hagino}, Kouichi and {Hamaguchi}, Kenji and {Hatsukade}, Isamu and {Hayashi}, Katsuhiro and {Hayashi}, Takayuki and {Hell}, Natalie and {Hodges-Kluck}, Edmund and {Horiuchi}, Takafumi and {Hornschemeier}, Ann and {Hoshino}, Akio and {Ichinohe}, Yuto and {Ikuta}, Chisato and {Iizuka}, Ryo and {Ishi}, Daiki and {Ishida}, Manabu and {Ishihama}, Naoki and {Ishikawa}, Kumi and {Ishimura}, Kosei and {Jaffe}, Tess and {Katsuda}, Satoru and {Kanemaru}, Yoshiaki and {Kenyon}, Steven and {Kilbourne}, Caroline and {Kimball}, Mark and {Kitamoto}, Shunji and {Kobayashi}, Shogo and {Kohmura}, Takayoshi and {Kubota}, Aya and {Leutenegger}, Maurice and {Maeda}, Yoshitomo and {Markevitch}, Maxim and {Matsumoto}, Hironori and {Matsuzaki}, Keiichi and {McCammon}, Dan and {McLaughlin}, Brian and {McNamara}, Brian and {Mernier}, Francois and {Miko}, Joseph and {Miller}, Jon and {Minesugi}, Kenji and {Mitani}, Shinji and {Mitsuishi}, Ikuyuki and {Mizumoto}, Misaki and {Mizuno}, Tsunefumi and {Mukai}, Koji and {Murakami}, Hiroshi and {Mushotzky}, Richard and {Nakazawa}, Kazuhiro and {Natsukari}, Chikara and {Ness}, Jan-Uwe and {Nigo}, Kenichiro and {Nishiyama}, Mari and {Nobukawa}, Kumiko and {Nobukawa}, Masayoshi and {Noda}, Hirofumi and {Odaka}, Hirokazu and {Ogawa}, Mina and {Ogawa}, Shoji and {Ogorzalek}, Anna and {Okajima}, Takashi and {Okamoto}, Atsushi and {Ota}, Naomi and {Ozaki}, Masanobu and {Paltani}, Stephane and {Plucinsky}, Paul and {Porter}, F. Scott and {Pottschmidt}, Katja and {Quero}, Jose Antonio and {Sasaki}, Takahiro and {Sato}, Kosuke and {Sato}, Rie and {Sato}, Toshiki and {Sato}, Yoichi and {Seta}, Hiromi and {Shida}, Maki and {Shidatsu}, Megumi and {Shigeto}, Shuhei and {Shipman}, Russel and {Shinozaki}, Keisuke and {Shirron}, Peter and {Simionescu}, Aurora and {Smith}, Randall and {Soong}, Yang and {Suzuki}, Hiromasa and {Szymkowiak}, Andrew and {Takahashi}, Hiromitsu and {Takeo}, Mai and {Tamagawa}, Toru and {Tamura}, Keisuke and {Tanaka}, Takaaki and {Tanimoto}, Atsushi and {Terashima}, Yuichi and {Tsuboi}, Yohko and {Tsujimoto}, Masahiro and {Tsunemi}, Hiroshi and {Tsuru}, Takeshi and {Uchida}, Hiroyuki and {Uchida}, Nagomi and {Uchida}, Yuusuke and {Uchiyama}, Hideki and {Uno}, Shinichiro and {Vink}, Jacco and {Witthoeft}, Michael and {Wolfs}, Rob and {Yamada}, Satoshi and {Yamada}, Shinya and {Yamaoka}, Kazutaka and {Yamasaki}, Noriko and {Yamauchi}, Makoto and {Yamauchi}, Shigeo and {Yanagase}, Keiichi and {Yaqoob}, Tahir and {Yasuda}, Susumu and {Yoneyama}, Tomokage and {Yoshida}, Tessei and {Yukita}, Miohoko},
        title = "{X-Ray Imaging and Spectroscopy Mission}",
      journal = {\pasj},
         year = 2025,
        month = apr,
          doi = {10.1093/pasj/psaf023},
       adsurl = {https://ui.adsabs.harvard.edu/abs/2025PASJ..tmp...28T}
}

@ARTICLE{vandermeulen24b,
       author = {{Vander Meulen}, Bert and {Camps}, Peter and {Savi{\'c}}, {\DH}or{\dj}e and {Baes}, Maarten and {Matt}, Giorgio and {Stalevski}, Marko},
        title = "{X-ray polarisation in AGN circumnuclear media: Polarisation framework and 2D torus models}",
      journal = {\aap},
         year = 2024,
        month = sep,
       volume = {689},
          eid = {A297},
        pages = {A297},
          doi = {10.1051/0004-6361/202450773},
archivePrefix = {arXiv},
       eprint = {2409.02986},
 primaryClass = {astro-ph.HE},
       adsurl = {https://ui.adsabs.harvard.edu/abs/2024A&A...689A.297V}
}

@article{yusefzadeh84,
	adsurl = {https://ui.adsabs.harvard.edu/abs/1984ApJ...278..186Y},
	author = {{Yusef-Zadeh}, F. and {Morris}, M. and {White}, R.~L.},
	doi = {10.1086/161780},
	journal = {\apj},
	month = mar,
	pages = {186-194},
	title = {{Bipolar reflection nebulae : Monte Carlo simulations.}},
	volume = {278},
	year = 1984}

@article{stalevski16,
	adsurl = {https://ui.adsabs.harvard.edu/abs/2016MNRAS.458.2288S},
	archiveprefix = {arXiv},
	author = {{Stalevski}, Marko and {Ricci}, Claudio and {Ueda}, Yoshihiro and {Lira}, Paulina and {Fritz}, Jacopo and {Baes}, Maarten},
	doi = {10.1093/mnras/stw444},
	eprint = {1602.06954},
	journal = {\mnras},
	month = may,
	number = {3},
	pages = {2288-2302},
	primaryclass = {astro-ph.GA},
	title = {{The dust covering factor in active galactic nuclei}},
	volume = {458},
	year = 2016}

@article{weisskopf22,
	adsurl = {https://ui.adsabs.harvard.edu/abs/2022JATIS...8b6002W},
	archiveprefix = {arXiv},
	author = {{Weisskopf}, Martin C. and {Soffitta}, Paolo and {Baldini}, Luca and {Ramsey}, Brian D. and {O'Dell}, Stephen L. and {Romani}, Roger W. and {Matt}, Giorgio and {Deininger}, William D. and {Baumgartner}, Wayne H. and {Bellazzini}, Ronaldo and {Costa}, Enrico and {Kolodziejczak}, Jeffery J. and {Latronico}, Luca and {Marshall}, Herman L. and {Muleri}, Fabio and {Bongiorno}, Stephen D. and {Tennant}, Allyn and {Bucciantini}, Niccolo and {Dovciak}, Michal and {Marin}, Frederic and {Marscher}, Alan and {Poutanen}, Juri and {Slane}, Pat and {Turolla}, Roberto and {Kalinowski}, William and {Di Marco}, Alessandro and {Fabiani}, Sergio and {Minuti}, Massimo and {La Monaca}, Fabio and {Pinchera}, Michele and {Rankin}, John and {Sgro'}, Carmelo and {Trois}, Alessio and {Xie}, Fei and {Alexander}, Cheryl and {Allen}, D. Zachery and {Amici}, Fabrizio and {Andersen}, Jason and {Antonelli}, Angelo and {Antoniak}, Spencer and {Attin{\`a}}, Primo and {Barbanera}, Mattia and {Bachetti}, Matteo and {Baggett}, Randy M. and {Bladt}, Jeff and {Brez}, Alessandro and {Bonino}, Raffaella and {Boree}, Christopher and {Borotto}, Fabio and {Breeding}, Shawn and {Brienza}, Daniele and {Bygott}, H. Kyle and {Caporale}, Ciro and {Cardelli}, Claudia and {Carpentiero}, Rita and {Castellano}, Simone and {Castronuovo}, Marco and {Cavalli}, Luca and {Cavazzuti}, Elisabetta and {Ceccanti}, Marco and {Centrone}, Mauro and {Citraro}, Saverio and {D'Amico}, Fabio and {D'Alba}, Elisa and {Di Gesu}, Laura and {Del Monte}, Ettore and {Dietz}, Kurtis L. and {Di Lalla}, Niccolo' and {Persio}, Giuseppe Di and {Dolan}, David and {Donnarumma}, Immacolata and {Evangelista}, Yuri and {Ferrant}, Kevin and {Ferrazzoli}, Riccardo and {Ferrie}, MacKenzie and {Footdale}, Joseph and {Forsyth}, Brent and {Foster}, Michelle and {Garelick}, Benjamin and {Gunji}, Shuichi and {Gurnee}, Eli and {Head}, Michael and {Hibbard}, Grant and {Johnson}, Samantha and {Kelly}, Erik and {Kilaru}, Kiranmayee and {Lefevre}, Carlo and {Roy}, Shelley Le and {Loffredo}, Pasqualino and {Lorenzi}, Paolo and {Lucchesi}, Leonardo and {Maddox}, Tyler and {Magazzu}, Guido and {Maldera}, Simone and {Manfreda}, Alberto and {Mangraviti}, Elio and {Marengo}, Marco and {Marrocchesi}, Alessandra and {Massaro}, Francesco and {Mauger}, David and {McCracken}, Jeffrey and {McEachen}, Michael and {Mize}, Rondal and {Mereu}, Paolo and {Mitchell}, Scott and {Mitsuishi}, Ikuyuki and {Morbidini}, Alfredo and {Mosti}, Federico and {Nasimi}, Hikmat and {Negri}, Barbara and {Negro}, Michela and {Nguyen}, Toan and {Nitschke}, Isaac and {Nuti}, Alessio and {Onizuka}, Mitch and {Oppedisano}, Chiara and {Orsini}, Leonardo and {Osborne}, Darren and {Pacheco}, Richard and {Paggi}, Alessandro and {Painter}, Will and {Pavelitz}, Steven D. and {Pentz}, Christina and {Piazzolla}, Raffaele and {Perri}, Matteo and {Pesce-Rollins}, Melissa and {Peterson}, Colin and {Pilia}, Maura and {Profeti}, Alessandro and {Puccetti}, Simonetta and {Ranganathan}, Jaganathan and {Ratheesh}, Ajay and {Reedy}, Lee and {Root}, Noah and {Rubini}, Alda and {Ruswick}, Stephanie and {Sanchez}, Javier and {Sarra}, Paolo and {Santoli}, Francesco and {Scalise}, Emanuele and {Sciortino}, Andrea and {Schroeder}, Christopher and {Seek}, Tim and {Sosdian}, Kalie and {Spandre}, Gloria and {Speegle}, Chet O. and {Tamagawa}, Toru and {Tardiola}, Marcello and {Tobia}, Antonino and {Thomas}, Nicholas E. and {Valerie}, Robert and {Vimercati}, Marco and {Walden}, Amy L. and {Weddendorf}, Bruce and {Wedmore}, Jeffrey and {Welch}, David and {Zanetti}, Davide and {Zanetti}, Francesco},
	doi = {10.1117/1.JATIS.8.2.026002},
	eid = {026002},
	eprint = {2112.01269},
	journal = {Journal of Astronomical Telescopes, Instruments, and Systems},
	month = apr,
	number = {2},
	pages = {026002},
	primaryclass = {astro-ph.IM},
	title = {{The Imaging X-Ray Polarimetry Explorer (IXPE): Pre-Launch}},
	volume = {8},
	year = 2022}

@article{camps15a,
	adsurl = {https://ui.adsabs.harvard.edu/abs/2015A&C.....9...20C},
	archiveprefix = {arXiv},
	author = {{Camps}, P. and {Baes}, M.},
	doi = {10.1016/j.ascom.2014.10.004},
	eprint = {1410.1629},
	journal = {Astronomy and Computing},
	month = mar,
	pages = {20-33},
	primaryclass = {astro-ph.IM},
	title = {{SKIRT: An advanced dust radiative transfer code with a user-friendly architecture}},
	volume = {9},
	year = 2015}

@article{camps20,
	adsurl = {https://ui.adsabs.harvard.edu/abs/2020A&C....3100381C},
	archiveprefix = {arXiv},
	author = {{Camps}, P. and {Baes}, M.},
	doi = {10.1016/j.ascom.2020.100381},
	eid = {100381},
	eprint = {2003.00721},
	journal = {Astronomy and Computing},
	month = apr,
	pages = {100381},
	primaryclass = {astro-ph.GA},
	title = {{SKIRT 9: Redesigning an advanced dust radiative transfer code to allow kinematics, line transfer and polarization by aligned dust grains}},
	volume = {31},
	year = 2020}

@article{peest17,
	adsurl = {https://ui.adsabs.harvard.edu/abs/2017A&A...601A..92P},
	archiveprefix = {arXiv},
	author = {{Peest}, C. and {Camps}, P. and {Stalevski}, M. and {Baes}, M. and {Siebenmorgen}, R.},
	doi = {10.1051/0004-6361/201630157},
	eid = {A92},
	eprint = {1702.07354},
	journal = {\aap},
	month = may,
	pages = {A92},
	primaryclass = {astro-ph.IM},
	title = {{Polarization in Monte Carlo radiative transfer and dust scattering polarization signatures of spiral galaxies}},
	volume = {601},
	year = 2017}

@article{vandermeulen23,
	adsurl = {https://ui.adsabs.harvard.edu/abs/2023A&A...674A.123V},
	archiveprefix = {arXiv},
	author = {{Vander Meulen}, Bert and {Camps}, Peter and {Stalevski}, Marko and {Baes}, Maarten},
	doi = {10.1051/0004-6361/202245783},
	eid = {A123},
	eprint = {2304.10563},
	journal = {\aap},
	month = jun,
	pages = {A123},
	primaryclass = {astro-ph.HE},
	title = {{X-ray radiative transfer in full 3D with SKIRT}},
	volume = {674},
	year = 2023}

@article{wilms00,
	adsurl = {https://ui.adsabs.harvard.edu/abs/2000ApJ...542..914W},
	archiveprefix = {arXiv},
	author = {{Wilms}, J. and {Allen}, A. and {McCray}, R.},
	doi = {10.1086/317016},
	eprint = {astro-ph/0008425},
	journal = {\apj},
	month = oct,
	number = {2},
	pages = {914-924},
	primaryclass = {astro-ph},
	title = {{On the Absorption of X-Rays in the Interstellar Medium}},
	volume = {542},
	year = 2000}

@ARTICLE{Balucinska-Church_2000a,
   author = {{Ba{\l}uci{\'n}ska-Church}, M. and {Church}, M.~J. and
                  {Charles}, P.~A. and {Nagase}, F. and {LaSala},
                  J. and {Barnard}, R.},
    title = "{The distribution of X-ray dips with orbital phase in Cygnus X-1}",
  journal = {\mnras},
   eprint = {arXiv:astro-ph/9909235},
     year = 2000,
    month = feb,
   volume = 311,
    pages = {861-868},
      doi = {10.1046/j.1365-8711.2000.03149.x},
   adsurl = {http://adsabs.harvard.edu/abs/2000MNRAS.311..861B}
}

@ARTICLE{Boroson_2010a,
   author = {{Boroson}, B. and {Vrtilek}, S.~D.},
    title = "{X-ray Variations at the Orbital Period from Cygnus X-1 IN the High/Soft State}",
  journal = {\apj},
archivePrefix = "arXiv",
   eprint = {0912.5412},
 primaryClass = "astro-ph.HE",
     year = 2010,
    month = feb,
   volume = 710,
    pages = {197-206},
      doi = {10.1088/0004-637X/710/1/197},
   adsurl = {http://adsabs.harvard.edu/abs/2010ApJ...710..197B}
}

@ARTICLE{Gies_2003a,
   author = {{Gies}, D.~R. and {Bolton}, C.~T. and {Thomson}, J.~R. and {Huang}, W. and 
{McSwain}, M.~V. and {Riddle}, R.~L. and {Wang}, Z. and {Wiita}, P.~J. and 
{Wingert}, D.~W. and {Cs{\'a}k}, B. and {Kiss}, L.~L.},
 title = "{Wind Accretion and State Transitions in Cygnus X-1}",
  journal = {\apj},
   eprint = {arXiv:astro-ph/0206253},
     year = 2003,
    month = jan,
   volume = 583,
    pages = {424-436},
      doi = {10.1086/345345},
   adsurl = {http://adsabs.harvard.edu/abs/2003ApJ...583..424G}
}

@ARTICLE{Sowers_1998a,
   author = {{Sowers}, J.~W. and {Gies}, D.~R. and {Bagnuolo}, W.~G. and 
	{Shafter}, A.~W. and {Wiemker}, R. and {Wiggs}, M.~S.},
    title = "{Tomographic Analysis of H{$\alpha$} Profiles in HDE 226868/Cygnus X-1}",
  journal = {\apj},
     year = 1998,
    month = oct,
   volume = 506,
    pages = {424-430},
      doi = {10.1086/306246},
   adsurl = {http://adsabs.harvard.edu/abs/1998ApJ...506..424S}
}

@ARTICLE{Hanke_2009a,
   author = {{Hanke}, M. and {Wilms}, J. and {Nowak}, M.~A. and {Pottschmidt}, K. and 
	{Schulz}, N.~S. and {Lee}, J.~C.},
    title = "{Chandra X-Ray Spectroscopy of the Focused Wind in the Cygnus X-1 System. I. The Nondip Spectrum in the Low/Hard State}",
  journal = {\apj},
archivePrefix = "arXiv",
   eprint = {0808.3771},
     year = 2009,
    month = jan,
   volume = 690,
    pages = {330-346},
      doi = {10.1088/0004-637X/690/1/330},
   adsurl = {http://adsabs.harvard.edu/abs/2009ApJ...690..330H}
}

@ARTICLE{El_Mellah_2019a,
       author = {{El Mellah}, I. and {Sander}, A.~A.~C. and {Sundqvist}, J.~O. and {Keppens}, R.},
        title = "{Formation of wind-captured disks in supergiant X-ray binaries. Consequences for Vela X-1 and Cygnus X-1}",
      journal = {\aap},
         year = 2019,
        month = feb,
       volume = {622},
          eid = {A189},
        pages = {A189},
          doi = {10.1051/0004-6361/201834498},
archivePrefix = {arXiv},
       eprint = {1810.12933},
 primaryClass = {astro-ph.HE},
       adsurl = {https://ui.adsabs.harvard.edu/abs/2019A&A...622A.189E}
}

@ARTICLE{Owocki_1984a,
   author = {{Owocki}, S.~P. and {Rybicki}, G.~B.},
    title = "{Instabilities in line-driven stellar winds. I - Dependence on perturbation wavelength}",
  journal = {\apj},
     year = 1984,
    month = sep,
   volume = 284,
    pages = {337-350},
      doi = {10.1086/162412},
   adsurl = {http://adsabs.harvard.edu/abs/1984ApJ...284..337O}
}

\end{document}